%% file: main.tex
\documentclass[%
preprint, superscriptaddress, amsmath,amssymb, aps, pra, showkeys]{revtex4-2}

\input{preamble/new_commands}
\input{preamble/header.tex}

\begin{document}

\title{Field output correction factors using a fully characterized plastic scintillation detector}

\author{Luc Gingras}
\affiliation{Service de physique médicale et de radioprotection, Centre intégré de cancérologie, CHU de Québec–Université Laval et Centre de recherche du CHU de Québec, Québec, Québec, Canada}

\author{Yunuen Cervantes}
\affiliation{Département de physique, de génie physique et d'optique, et Centre de recherche sur le cancer, Université Laval, Québec, Québec, Canada}
\affiliation{Service de physique médicale et de radioprotection, Centre intégré de cancérologie, CHU de Québec–Université Laval et Centre de recherche du CHU de Québec, Québec, Québec, Canada}

\author{Frederic Beaulieu}
\affiliation{Service de physique médicale et de radioprotection, Centre intégré de cancérologie, CHU de Québec–Université Laval et Centre de recherche du CHU de Québec, Québec, Québec, Canada}

\author{Magali Besnier}
\affiliation{Service de physique médicale et de radioprotection, Centre intégré de cancérologie, CHU de Québec–Université Laval et Centre de recherche du CHU de Québec, Québec, Québec, Canada}

\author{Benjamin Coté}
\affiliation{Medscint Inc., Québec, QC, Canada}

\author{Simon Lambert-Girard}
\affiliation{Medscint Inc., Québec, QC, Canada}

\author{Danahé LeBlanc}
\affiliation{Medscint Inc., Québec, QC, Canada}

\author{Yoan LeChasseur}
\affiliation{Medscint Inc., Québec, QC, Canada}

\author{François Therriault-Proulx}
\affiliation{Medscint Inc., Québec, QC, Canada}

\author{Luc Beaulieu}
\affiliation{Département de physique, de génie physique et d'optique, et Centre de recherche sur le cancer, Université Laval, Québec, Québec, Canada}
\affiliation{Service de physique médicale et de radioprotection, Centre intégré de cancérologie, CHU de Québec–Université Laval et Centre de recherche du CHU de Québec, Québec, Québec, Canada}

\author{Louis Archambault}
\email{louis.archambault@phy.ulaval.ca}
\affiliation{Département de physique, de génie physique et d'optique, et Centre de recherche sur le cancer, Université Laval, Québec, Québec, Canada}
\affiliation{Service de physique médicale et de radioprotection, Centre intégré de cancérologie, CHU de Québec–Université Laval et Centre de recherche du CHU de Québec, Québec, Québec, Canada}

\date{\today}

\begin{abstract}
\noindent \textbf{Introduction}: As small radiation fields play an ever-increasing
role in radiation therapy, accurate dosimetry of these fields becomes critical to
ensure high quality dose calculation and treatment optimization. Despite the
availability of several small volume dose detectors, small field dosimetry
remains challenging. The \probe{}, a new plastic scintillation detector (PSD) part of the Hyperscint RP-200 dosimetric platform from Medscint, that requires only minimal corrections can potentially facilitate small field measurements. 

In this work our objective is twofold: first we performed a detailed
characterization of the field output correction factors of the \probe{} PSD over
a wide range of field sizes; second we demonstrate how this PSD can be used to
determine the field output correction factors for other small field detectors. In
addition, we carefully studied uncertainties in order to provide a detailed
uncertainty budget for a wide range of dosimeters.

\noindent \textbf{Methods}: Our work is based on IAEA TRS-483 report. EGSnrc Monte Carlo
simulations of the \probe{} were conducted to determine the impact of
detector composition, surrounding materials, dose averaging within the sensitive
volume as well as ionization quenching. From these simulations, the field output
correction factors of this PSD were determined. Then, by experimental comparisons, field output correction factors for 2 solid state detectors and 3 small
volume ion chambers have been determined.

\noindent \textbf{Results}: With \probe{}, the material composition factor is
well balanced with the ionization quenching making the field output correction
factor near unity. For fields between 0.6~$\times$~0.6 and
30~$\times$~30~cm$^2$, the field output correction factors of the \probe{} were
between 1.002 and 0.999 with a total uncertainty of 0.5~\%. Analysis of the
uncertainty budget showed that, using \probe{} for measuring output factors an
overall uncertainty of 0.59\% can be achieved for a 1~$\times$~1~cm$^2$ field
size. 

\noindent \textbf{Conclusion}: With field output correction factors close to
unity for a wide range of field sizes, the \probe{} PSD is a near-ideal detector
for small field dosimetry. Furthermore, it can be used to experimentally
determine the field output correction factors of other dosimeters with great
accuracy.

\end{abstract}


\maketitle


\input{content/introduction}

\input{content/methods_and_materials}

\input{content/results}

\input{content/discussion}
\input{content/conclusion}

\input{content/acknowledgements}

\bibliographystyle{medphy.bst}
\bibliography{small_field}

\pagebreak
\appendix 

\input{content/appendix/k_defs.tex}

\input{content/appendix/det_geo-orient_integration.tex}
\input{content/appendix/kvol_psd_all_profiles.tex}
\input{content/appendix/kfclin_comp.tex}

\end{document}

%% file: preamble/new_commands.tex
\newcommand{\kfclin}{k_{Q_{\mathrm{clin}} ,Q_{\mathrm{msr}}}^{f_{\mathrm{clin}} ,f_{\mathrm{msr}}}}

\newcommand{\OF}{\Omega_{Q_{\mathrm{clin}} ,Q_{\mathrm{msr}}}^{f_{\mathrm{clin}} ,f_{\mathrm{msr}}}}

\newcommand{\SCLIN}{S_{\mathrm{clin}}}

\newcommand{\pwall}{P_{\mathrm{wall}}}

\newcommand{\pscint}{P_{\mathrm{scint}}}

\newcommand{\kvol}{k_{\mathrm{vol}}}

\newcommand{\kioq}{k_{\mathrm{ioq}}}

\newcommand{\kpos}{k_{\mathrm{pos}}}

\newcommand{\kpol}{k_{\mathrm{pol}}}
\newcommand{\kion}{k_{\mathrm{ion}}}

\newcommand{\probe}{PRB-0002} 

\newcommand{\Mratio}{M_{Q_{\mathrm{clin}},Q_{\mathrm{ref}}}^{f_{\mathrm{clin}},f_{\mathrm{ref}}}}

%% file: preamble/header.tex
\usepackage{graphicx}
\usepackage{dcolumn}
\usepackage{colortbl}
\usepackage{xcolor}
\usepackage{siunitx}
\usepackage{amsmath}
\usepackage{multirow}
\usepackage{booktabs}
\usepackage{xspace} 
\usepackage{physics}


%% file: content/introduction.tex
\section{Introduction}

Radiation therapy (RT) has evolved from treating large anatomical areas
encompassing both the cancer and copious amount of healthy tissues to delivering
dose distributions that can precisely match the shape of a tumor. This evolution
has been driven by the increased mechanical capabilities of medical linear
accelerators (linacs) and by the widespread availability of imaging
modalities (MRI, CT, CBCT) both prior and during treatment. The underlying
hypothesis behind modern RT is that treating smaller volumes makes it possible
to maximize the dose to the cancer while minimizing damage to healthy tissues thus
achieving better clinical outcomes. 

Treating small volumes necessarily involves at least some small radiation fields.
Indeed, most advanced treatment techniques such as intensity modulated radiation
therapy (IMRT), volumetric modulated arc therapy (VMAT) and stereotactic
radiosurgery (SRS) rely heavily on small fields. To guarantee the accuracy of
these treatments, it is of a prime importance that small fields are well
characterized and well modeled in treatment planning systems. As pointed out by
Lechner et al.~\cite{lechner-2018}: ``\emph{the accuracy of small field
OFs directly impacts the accuracy of patient dose calculation.}''
 
Dosimetry of small fields has been studied for decades either directly or
through the broader categories of non-equilibrium~\cite{das-2008} or
non-standard~\cite{alfonso.2008} fields. Today, the most accurate definition is
provided by the IAEA-AAPM TRS-483 code of practice~\cite{international2017iaea}:
a field is considered small if at least one of the following conditions is met:
(1)~a loss of lateral charged particle equilibrium on the beam axis; (2)~a
partial occlusion of the primary photon source by the collimating device on the
beam axis; (3)~the size of the detector is similar or larger than the beam
dimensions. While the conditions 1 and 2 are directly related to the beam,
condition 3 depends on the detector used to perform a measurement. Thus, a field
might be considered small for one detector and not for another.

The challenges of small field dosimetry are well documented
(see~\cite{alfonso.2008,international2017iaea} for a good summary). According to
TRS-483, the increased usage of small fields ``\emph{has increased the
uncertainty of clinical dosimetry and weakened its traceability to reference
dosimetry based on conventional [codes of
practice]}''~\cite{international2017iaea}. It is in this context that the
IAEA-AAPM proposed the TRS-483 code of practice to provide a standardized method
for the determination of field output factors for relative dosimetry in small
static photon fields.

A field output factor, $\Omega$, is the ratio of the dose to water at the center
of two static fields: a clinical field, $f_\text{clin}$ (of quality
$Q_\text{clin}$) and a reference field, either a standard reference,
$f_\text{ref}$ (of quality $Q$), or a machine specific reference, $f_\text{msr}$
(of quality $Q_\text{msr}$). For large field sizes, the dose ratio is well
approximated by the ratio of measured dosimeter signals, because the conversion
from a detector measurement to a dose to water is mostly independent of field
size. However, for small fields, this conversion depends on the field size
because of the volume averaging, the perturbation of the charged particle
fluence due to the presence of the detector and the intrinsic change in beam
quality with field size. A field output correction factor, $\kfclin$ must
therefore be applied to the ratio of measurements in order to determine $\Omega$
of small fields. 

Accurate determination of $\kfclin$ is one of the main challenge of small field
dosimetry. Values are tabulated in TRS-483~\cite{international2017iaea} for a
wide range of detectors. This data, taken from the literature, consider both
experimental measurements and Monte Carlo (MC) simulations. Nevertheless, it
might be necessary to determine $\kfclin$ for detectors or beam quality not
listed~\cite{Das.2021bms, Casar.2019, mcgrath-2022} or at higher accuracy level,
for detectors in different orientations~\cite{Casar.2020} or to perform an
independent evaluation of tabulated values. In their initial report, Alfonso et
al.~\cite{alfonso.2008} mentioned calculating this factor with MC simulation
alone or measuring it with a \emph{suitable}, well characterized detector. 

The search for an ideal reference detector for small field dosimetry is ongoing.
The list of properties for such a detector is well known (see Table~6 of
\cite{international2017iaea}). Some examples include: stability, spatial
resolution, dose and dose rate linearity. Also, in general, a good reference
detector should have an $\kfclin$ as close to 1 as possible. Because of their
large correction factors, ion chambers (including small-volume ion chambers) are
usually not ideal for reference dosimetry of small fields, despite being the
gold standard for other types of beams. Diodes, diamond detector, radiochromic
films, plastic scintillation detectors (PSDs) and alanine dosimeters have all
been investigated as potential candidates for reference measurements, but none
of them has emerged as clear choice.

Plastic scintillation detectors possess many advantages for small field
dosimetry \cite{gagnon-2011, Papaconstadopoulos.2017} and were initially thought
to have an output correction factor close to unity~\cite{international2017iaea}.
Because of these properties, Casar et al.~\cite{Casar.2019} have recently used a
PSD as a reference detector to determine the $\kfclin$ of several other detectors. 

In a PSD, the scintillation signal is usually assumed to be proportional to the
dose received~\cite{beddar-1992a} although PSDs are known to suffer from
ionization quenching (i.e. a diminution of the scintillation light emitted per
unit dose) when irradiated by beams of particles of high linear energy transfer
like protons and other hadrons~\cite{Wang.2011}. It has been recently shown by
Santurio et al.~\cite{Santurio.2019, Santurio.2020} that ionization quenching
can also be observed in megavoltage photon beams because of the presence of low
energy secondary electrons. This means that the signal produced by a PSD can
depend on beam quality and field size. To account for this, the ionization
quenching correction factor, $k_\text{ioq}$ was introduced~\cite{Santurio.2019}
and determined using Monte Carlo simulations. Including the effect of ionization
quenching explicitly in $\kfclin$ implies that the output correction factor
will differ from unity, contrary to what has been assumed so far.

In this context, the new Medscint's \probe{} PSD has been developed by Medscint,
(Medscint inc., Quebec city, Canada) and previously
characterized~\cite{Uijtewaal.2023}. The scintillator probe, a component of the
Hyperscint RP-200 dosimetry Platform, was designed to have a $\kfclin$ as close
to 1 as possible even when including the impact of $k_\text{ioq}$. The goal of
this work is (1) to determine the output correction factor of this new probe
using Monte Carlo simulations and then (2) to use it as a reference in order to
determine the $\kfclin$ of other detectors for small fields.

%% file: content/methods_and_materials.tex
\section{Methods and Materials}

\input{content/method/meth_form_and_def.tex}

\input{content/method/meth_corr.tex}

\input{content/method/meth_kfclin_hyperscint.tex}

\input{content/method/meth_detectors.tex}

\input{content/method/meth_measurements.tex}

\input{content/method/meth_uncertainties.tex}

\input{content/method/meth_kfclin_all.tex}

%% file: content/method/meth_form_and_def.tex
\subsection{Formalism and definitions} \label{sec:formalism}

\input{content/method/meth_formalism.tex}


While the signal of an ideal dosimeter should only be proportional to the
average dose deposited in its sensitive volume,
($\overline{D}_{\mathrm{det},Q}^{f}$), most detector exhibit some dependency to
other factors. The signal production per unit dose may depend on environmental
factors, operating conditions, beam quality, as well as temporal dynamics of dose
delivery. Signal collection can also be affected by operating conditions and
detector-reader coupling configuration. Factors affecting signal production and
collection are discussed in length in~\cite{Almond.1999, international2017iaea}

We distinguish two types of correction factors that can be applied to raw
measurement signals ($M_\mathrm{raw}$) of a detector to transform it into a quantity
solely dependent on the dose absorbed by its sensitive volume. The first type
(I) includes corrections for well known influence quantities (e.g. temperature,
pressure) that are independent of small changes in beam quality. In other word,
the value of a type~I correction factor should not change when going from a beam
of quality $Q_\mathrm{msr}$ to a beam of quality $Q_\mathrm{clin}$. The second
type (II) includes corrections for influence quantities that may vary with
small changes in beam quality. This can be represented by the following
equations:
\begin{align}
    \overline{D}_{\mathrm{det},Q}^{f} &\propto (M_\mathrm{raw})_{Q}^{f} \times \prod_{i} (k_{I,i})_{Q}^{f} \times \prod_{j}(k_{II,j})_{Q}^{f},  \label{eq:m_det1} \\
    \overline{D}_{\mathrm{det},Q}^{f} &\propto M_{Q}^{f} \times \prod_{j}(k_{II,j})_{Q}^{f},  \label{eq:m_det2}
\end{align}
where $(M_{raw})_{Q}^{f}$ and $M_{Q}^{f}$ are respectively the raw measurement
signal and the measurement signal corrected for all influence quantities
independant of small changes in beam quality. \emph{Small changes} in beam
quality are assumed to be changes that can easily occur in a measurement
sequence. For example, changes in $Q$ caused by varying the field size of a
photon beam. From equation~(\ref{eq:m_det2}), we can express
the ratio of measurements corrected for type~I factors:
\begin{equation}
    \frac{M_{Q_\mathrm{clin}}^{f_\mathrm{clin}}}{M_{Q_\mathrm{msr}}^{f_\mathrm{msr}}} 
    = \left[ \frac{\overline{D}_{\mathrm{det},Q_\mathrm{clin}}^{f_\mathrm{clin}}}{\overline{D}_{\mathrm{det},Q_\mathrm{msr}}^{f_\mathrm{msr}}} \right] 
    \times \left[ \frac{1}{\prod_{j}(k_{II,j})_{Q_\mathrm{clin},Q_\mathrm{msr}}^{f_\mathrm{clin},f_\mathrm{msr}}} \right], \label{eq:m_det_ratio}
\end{equation}
Combining eq.~(\ref{eq:m_det_ratio}) and eq.~(\ref{eq:kfclin}):
\begin{equation} \label{eq:kfclin2}
    k_{Q_{\mathrm{clin}} ,Q_{\mathrm{msr}}}^{f_{\mathrm{clin}} ,f_{\mathrm{msr}}} = \left[ \frac{\frac{D_{\mathrm{w},Q_{\mathrm{clin}}}^{f_{\mathrm{clin}}}}{D_{\mathrm{w},Q_{\mathrm{msr}}}^{f_{\mathrm{msr}}}}}{\frac{\overline{D}_{det,Q_{\mathrm{clin}}}^{f_{\mathrm{clin}}}}{\overline{D}_{det,Q_{\mathrm{msr}}}^{f_{\mathrm{msr}}}}} \right] \times \left[ \prod_{j}(k_{II,j})_{Q_\mathrm{clin},Q_\mathrm{msr}}^{f_\mathrm{clin},f_\mathrm{msr}} \right],
\end{equation}
Thus, $\kfclin$ determination should take into account differences in type~II
factors between $Q_\mathrm{clin}$ and $Q_\mathrm{msr}$ using
eq.~(\ref{eq:m_det_ratio}). Most Monte Carlo studies aiming to determine
$\kfclin$ will implicitly assume type~II corrections to be unity. While this
is reasonable for ionization chambers and solid state detectors in normal
clinical conditions, it is not generally true for all detectors and conditions.
For example, ionization quenching in PSDs should be accounted for by type~II
corrections. To compare $\kfclin$ values extracted from different studies, it is
imperative to make sure that all non negligible type~I measurement corrections
have been applied in the same way on experimental data and that Monte Carlo
$\kfclin$ estimations include type~II corrections if needed. 

In this work, all experimental measurements and Monte Carlo simulations were
performed on a conventional linear accelerator (linac) capable of delivering a
reference ($\mathrm{ref}$) $10\times10$~cm$^2$ field. Thus, in this case the
machine specific reference~($\mathrm{msr}$) and $\mathrm{ref}$ are identical
conditions.

%% file: content/method/meth_formalism.tex
Following TG-51 report~\cite{Almond.1999}, the absorbed dose to water of a field
$f$ of quality $Q$, noted $D_{w,Q}^{f}$, can be calculated from a detector
reading corrected for environmental conditions, signal collection effects and
positional uncertainty, and which is noted $M_{Q}^{f}$. For dosimetry with an
ion chamber detector in a clinical field ($\mathrm{clin}$)~\cite{alfonso.2008},
the dose to water in the absence of the detector is given by:
\begin{equation} \label{eq:Dose_fclin}
   D_{\mathrm{w}, Q_{\mathrm{clin}}}^{f_{\mathrm{clin}}} 
   = M_{Q_{\mathrm{clin}}}^{f_{\mathrm{clin}}} N_{D, \mathrm{w}, Q_0}^{f_{\mathrm{ref}}} 
   k_{Q_{\mathrm{msr}}, Q_0}^{f_{\mathrm{msr}}, f_{\mathrm{ref}}} \kfclin,
\end{equation}
where the reference class ionization chamber calibration coefficient
$N_{D,w,Q_0}^{f_\mathrm{ref}}$ is available from a standards laboratory's
reference beam of quality $Q_0$ in the conventional reference field
$f_\mathrm{ref}$. The measured detector reading for a given clinical field,
$M_{Q_\mathrm{clin}}^{f_\mathrm{clin}}$ is corrected first by a factor $\kfclin$
to account for the difference in beam quality between $(f_\mathrm{clin},
Q_\mathrm{clin})$ and the machine specific reference $(f_\mathrm{msr},
Q_\mathrm{msr})$ and then by
a factor $k_{Q_\mathrm{msr}, Q_0}^{f_\mathrm{msr}, f_\mathrm{ref}}$, to account
for the difference in beam quality between the reference condition,
$(f_\mathrm{ref},Q_0)$ and the machine specific reference. 

Based on the formalism of Alfonso et al.~\cite{alfonso.2008}, the field output
factor $\OF$ can be determined using equations~(17) and~(18) of
TRS-483~\cite{international2017iaea}:
\begin{equation} 
    \OF = \frac{D_{\mathrm{w},Q_{\mathrm{clin}}}^{f_{\mathrm{clin}}}}{D_{\mathrm{w},Q_{\mathrm{msr}}}^{f_{\mathrm{msr}}}} = \frac{M_{Q_{\mathrm{clin}}}^{f_{\mathrm{clin}}}}{M_{Q_{\mathrm{msr}}}^{f_{\mathrm{msr}}}} \kfclin, \label{eq:OF_TRS} 
\end{equation}
In order to determine the field output factor from a measurement ratio, it is
necessary to know the detector specific output correction factors, $\kfclin$. From eq.~(\ref{eq:OF_TRS}), we can write: 
\begin{equation} 
   \kfclin = \frac
      {\left(D_{\mathrm{w},Q_{\mathrm{clin}}}^{f_{\mathrm{clin}}}/D_{\mathrm{w},Q_{\mathrm{msr}}}^{f_{\mathrm{msr}}}\right)}
      {\left(M_{Q_{\mathrm{clin}}}^{f_{\mathrm{clin}}}/M_{Q_{\mathrm{msr}}}^{f_{\mathrm{msr}}}\right)},\label{eq:kfclin}
\end{equation}

As detailed in Appendix~II of~\cite{international2017iaea}, this correction
factor can be obtained by three approaches:  (\emph{i})~using a perturbation
free (except for volume averaging) reference detector to obtain the dose ratio
in equation~(\ref{eq:kfclin}); (\emph{ii})~using a reference detector with known
output correction factor to evaluate the same ratio; and (\emph{iii})~using
Monte Carlo simulations to directly determine equation~(\ref{eq:kfclin}).

The first method has been used with ``perturbation free'' reference detectors
such as alanine, TLDs, organic scintillators and radiochromic films
\cite{Azangwe.2014, Casar.2019}. While the second method can be used with
a well characterized detector used in the same irradiation condition as the studied detector \cite{Morin.2013, Casar.2019}. Assuming that the output correction factor is
known for detector~1 ($\mathrm{det1}$), this correction can be determined for
detector~2 ($\mathrm{det2}$) using eq.~(\ref{eq:OF_TRS}):
\begin{equation} \label{eq:kfclin_det2}
   \left[k_{Q_{\mathrm{clin}} ,Q_{\mathrm{msr}}}^{f_{\mathrm{clin}} ,f_{\mathrm{msr}}}\right]_{\mathrm{det}_2} = \left[ \frac{M_{Q_{\mathrm{clin}}}^{f_{\mathrm{clin}}}}{M_{Q_{\mathrm{msr}}}^{f_{\mathrm{msr}}}} k_{Q_{\mathrm{clin}} ,Q_{\mathrm{msr}}}^{f_{\mathrm{clin}} ,f_{\mathrm{msr}}}\right]_{\mathrm{det}_1} \left[\frac{M_{Q_{\mathrm{msr}}}^{f_{\mathrm{msr}}}}{M_{Q_{\mathrm{clin}}}^{f_{\mathrm{clin}}}} \right]_{\mathrm{det}_2}.
\end{equation}

The third method involves the use of Monte Carlo calculations simulating both
the radiation source with its collimation device and the irradiated phantom
medium. Simulation material can be either water for the $D^f_Q$ ratio or
complete detector material and geometry in water for the $M^f_Q$ ratio in eq.~\ref{eq:kfclin}.
A hybrid method that uses MC simulations for $D^f_Q$ ratio and detector
measurements for $M^f_Q$ ratio has also been
proposed~\cite{Larraga-Gutierrez.2015}. However, such hybrid approach is
sensitive to the accuracy of radiation source model because particle energy
fluence distribution may be different between the simulation and measurement.

%% file: content/method/meth_corr.tex
\subsection{Measurement corrections} \label{sec:meas_corr}

For ionization chamber (IC), solid state (SS) and plastic scintillation
detectors (PSD), the type~I measurement correction equations take the
following forms:
\begin{align}
    M^\mathrm{IC} &= M^\mathrm{IC}_\mathrm{raw} k_\mathrm{TP} k_\mathrm{H} k_\mathrm{elec} k_\mathrm{pol} k_\mathrm{ion} k_\mathrm{drift} k_\mathrm{bg} k_\mathrm{stem} k_\mathrm{pos} \label{eq:mcorr_ic} \\
    M^\mathrm{SS} &= M^\mathrm{SS}_\mathrm{raw} k_\mathrm{T} k_\mathrm{elec} k_\mathrm{ion} k_\mathrm{drift} k_\mathrm{bg} k_\mathrm{stem} k_\mathrm{pos} \label{eq:mcorr_ss} \\
    M^\mathrm{PSD} &= M^{\mathrm{PSD}}_\mathrm{raw} k_\mathrm{T} k_\mathrm{read} k_\mathrm{drift} k_\mathrm{bg} k_\mathrm{stem} k_\mathrm{pos} \label{eq:mcorr_psd}
\end{align}
Most of these factors are well established in the literature; summarized
definitions can be found in Supplementary material~\ref{ap:kdef}

All measurements conducted in this study are relative measurements between a
$\mathrm{clin}$ field and a~$\mathrm{msr}$ field. Thus, the only factors needed
to be considered are those that change with field size or otherwise vary between
these two measurement conditions. Reference field measurements were repeated
frequently to ensure that drifts were small and could be taken into account when
performing measurement ratios. Type I correction factor ratios,
$(k_{I,i})^{f_\mathrm{clin},f_\mathrm{msr}}$, involving temperature, pressure,
humidity, readout/electrometer and machine output drifts were taken as unity.
Appropriate background subtraction also allowed to use $k_{bg}=1$. Furthermore,
$k_{stem}^{f_\mathrm{clin},f_\mathrm{msr}}$ ratios were taken as unity, by
selecting an appropriate field/detector orientation geometry. Uncertainties on
ratios assumed to be unity were accounted for, but without a systematic
component. Polarity, ionization recombination as well as positional variations
were all considered because of their field size dependence. For ionization
chambers, $k_\mathrm{pol}$ was determined for each field size by averaging
absolute positive and negative polarity raw measurements and dividing by the raw
measurement at the normal operating polarity. After each polarity reversal,
enough time and a pre-irradiation of at least 8~Gy were applied to guarantee a
stable detector output. 

\input{content/method/meth_k_ion.tex}

For ion chambers, these $k_\mathrm{ion}$ used for fitting were calculated by the
Boag two-voltage method \cite{Almond.1999} (validated by Jaffe plot regressions
\cite{Kry2012}) for a modified $\mathrm{msr}$ field ($\mathrm{msr}'$) at
source-to-surface distances (SSD) of 80, 90, 100 and 110~cm, and a constant
detector depth of 10~cm. Field sizes, $f_{\mathrm{msr}'}$, were determined by
scaling $f_\mathrm{msr}$ in order to maintain a constant effective field size at
the detector plane: $f_\mathrm{msr'} = f_\mathrm{msr} \times
100/(\mathrm{SSD}+10)$. This choice was made in order: (\emph{i}) to keep a
unity $k_\mathrm{stem}$ factor ratio by minimizing changes in stem irradiation
geometry, and (\emph{ii}) to provide an almost equivalent phantom scatter region
therefore making it possible to assume that both setups have the same effective
beam quality. $k_\mathrm{pol}$ values were measured for all conditions and
$k_\mathrm{pos}$ was taken as unity given the large field sizes and small
positioning error. 

For solid state detectors, even if the mechanisms involved in the dose per pulse
detector signal response dependence are fundamentally different than for ion
chambers, it is recognized that the simple mathematical form used in the TG-51
addendum for $k_\mathrm{ion}$ corrections of ion chambers can empirically be
applied also to solid state detector dose per pulse response dependence, at
least in the dose per pulse variation range encountered in standard radiation
therapy. Therefore, $k_\mathrm{ion}$ from equation~(\ref{eq:kion_ratio2}) will
also be used for correction of solid state detectors dose per pulse response
dependence. For these $SS$ detectors, $k_\mathrm{ion}$ ratios used for fitting
$B$ in eq.~(\ref{eq:kion_ratio2}) were obtained from the scaled $k_\mathrm{ion}$
ratios of the corresponding $\mathrm{msr}'$ geometry measured with an ion
chamber. This is motivated by the hypothesis that:
\begin{equation} \label{eq:kqssic_approx}
    \left[ k_{Q_{\mathrm{msr}'},Q_\mathrm{msr}}^{f_{\mathrm{msr}'},f_\mathrm{msr}} \right]_\mathrm{SS} \approx \left[ k_{Q_{\mathrm{msr}'},Q_\mathrm{msr}}^{f_{\mathrm{msr}'},f_\mathrm{msr}} \right]_\mathrm{IC} \approx 1,
\end{equation}
which is based on the facts that field sizes $f_{\mathrm{msr}'}$ and
$f_\mathrm{msr}$ are large relative to the detector sizes and that effective
beam quality at point of measurement remains constant at different SSD setups.
Therefore, using equations~(\ref{eq:kfclin_det2}), (\ref{eq:mcorr_ic}) and
(\ref{eq:mcorr_ss}), $k_\mathrm{ion}$ ratios of SS detectors for $\mathrm{msr}$
to $\mathrm{msr}'$ geometries can be approximated as:
\begin{equation} \label{eq:kion_ss}
    \left[ (k_\mathrm{ion})_{Q_{\mathrm{msr}'},Q_\mathrm{msr}}^{f_{\mathrm{msr}'},f_\mathrm{msr}} \right]_{SS} 
    \approx \frac{(M_\mathrm{raw}^\mathrm{IC})_{Q_{\mathrm{msr}'},Q_\mathrm{msr}}^{f_{\mathrm{msr}'},f_\mathrm{msr}}}{ (M_\mathrm{raw}^\mathrm{SS})_{Q_{\mathrm{msr}'},Q_\mathrm{msr}}^{f_{\mathrm{msr}'},f_\mathrm{msr}}} 
    \left[ (k_\mathrm{ion}k_\mathrm{pol})_{Q_{\mathrm{msr}'},Q_\mathrm{msr}}^{f_{\mathrm{msr}'},f_\mathrm{msr}} \right]_\mathrm{IC},
\end{equation}

Using these sets of $k_\mathrm{ion}$ ratios, $B$ can be determined for IC and SS
by inverting eq.~(\ref{eq:kion_ratio2}) for each $\mathrm{msr}'$. A single $B$
value for each detector was then obtained by averaging the result of $B$ for
$\mathrm{msr}'$ at SSDs of 80~cm and 110~cm. The uncertainty on $B$ was
estimated as the ratio of the maximal variation observed among all SSDs to the
mean value. With $B$, eq.~(\ref{eq:kion_ratio2}) was used to correct
ion-recombination effects of IC and SS detectors   

%% file: content/method/meth_k_ion.tex
$k_\mathrm{ion}$ correction factors for ionization chambers were determined using the TG-51
addendum~\cite{Mcewen2014} including a constant initial recombination term and a
detector dose per pulse dependant general recombination term. Using a
development similar to that of Duchaine et al.~\cite{Duchaine.2022}, it is
possible to obtain an expression for $k_\mathrm{ion}$ ratio:
\begin{equation} \label{eq:kion_ratio2}
   (k_\mathrm{ion})_{Q_\mathrm{clin},Q_\mathrm{msr}}^{f_\mathrm{clin},f_\mathrm{msr}} = \frac{1}{1 + B \left( 1 - (M_\mathrm{raw} k_\mathrm{pol} k_\mathrm{pos})_{Q_\mathrm{clin},Q_\mathrm{msr}}^{f_\mathrm{clin},f_\mathrm{msr}} \right)}.
\end{equation}
where $B$ is a parameter that depends on the dose per pulse in $\mathrm{msr}$
conditions. The value of $B$ was obtained by fitting eq.~(\ref{eq:kion_ratio2})
for IC and SS detectors using experimentally determined sets of
$k_\mathrm{ion}$.

%% file: content/method/meth_kfclin_hyperscint.tex
\subsection{Determination of $\kfclin$ for \probe{}}\label{sec:kfclin_psd}

The field output correction factors can be derived from a chain technique of
perturbation factors with respect to the dose to a point in water for a beam of
quality $Q$ and field size $f$, $D_{w,Q}^f$~\cite{Bouchard.2009}. For a PSD, the
chain is chosen based on the work of Papaconstadopoulos et
al.~\cite{Papaconstadopoulos.2014} with the additional correction factor
accounting for the ionization quenching of the scintillator, $\kioq$, a type~II
factor of influence. The perturbation chain entails four distinct geometries.
The first geometry corresponds to the fully assembled PSD, including the
protective jacket and clear optical fiber, in water. The second geometry is the
bare sensitive volume made of plastic scintillator in water. The third geometry
mimics the second, but is entirely composed of water. Finally, the last geometry
represents a point in water. Equation \ref{eq:chain} shows each step of the perturbation
chain:
\begin{align}
    D_{w,Q}^{f} &= (\kvol)_{Q}^{f} \cdot \overline{D}_{\mathrm{scint}(w),Q}^{f} \nonumber\\
    \overline{D}_{\mathrm{scint}(w),Q}^{f} &= \frac{\overline{D}_{\mathrm{scint}(w),Q}^{f}}{\overline{D}_{\mathrm{scint},Q}^{f}}\cdot \overline{D}_{\mathrm{scint},Q}^{f}
    = (\pscint)_{Q}^{f} \cdot \overline{D}_{\mathrm{scint},Q}^{f} \nonumber\\
    \overline{D}_{\mathrm{scint},Q}^{f} &= \frac{\overline{D}_{\mathrm{scint},Q}^{f}}{\overline{D}_{\mathrm{det},Q}^{f}} \cdot \overline{D}_{\mathrm{det},Q}^{f} = (\pwall)_{Q}^{f} \cdot \overline{D}_{\mathrm{det},Q}^{f}\label{eq:chain}
\end{align}
where $\overline{D}_{\mathrm{scint}(w),Q}^{f}$ is the average dose in the PSD
sensitive volume geometry composed of water, $\overline{D}_{\mathrm{scint},
Q}^{f}$ is the average dose to the PSD sensitive volume when scintillator
material composition is taken into account, and
$\overline{D}_{\mathrm{det},Q}^{f}$ is the average dose to the sensitive PSD
volume when considering the complete detector geometry. Collapsing all terms of
eq.~(\ref{eq:chain}) leads to:
\begin{equation} \label{eq:chain_coll}
    D_{w,Q}^{f} = (\kvol)_{Q}^{f} \cdot (\pscint)_{Q}^{f} \cdot (\pwall)_{Q}^{f} 
    \cdot \overline{D}_{det,Q}^{f}.
\end{equation}
$\pwall$ accounts for the impact of the PSD's wall, protective jacket and
optical fiber. $\pscint$ account for the perturbations due to the variation in
density and atomic composition of the scintillating material compared to water.
$\kvol$ is the volume averaging correction factor, its determination method is
described in section~\ref{sec:kvol}. 

Using eq.~(\ref{eq:chain_coll}) in eq.~(\ref{eq:kfclin2}) and assuming
ionization quenching is the only $k_{II}$ correction factor, the output
correction factor, $\kfclin$, for PSDs take the form:
\begin{equation} \label{eq:kfclin_psd}
   \kfclin = \left( \kvol \pscint \pwall \kioq \right)_{Q_\mathrm{clin},Q_\mathrm{msr}}^{f_\mathrm{clin},f_\mathrm{msr}}
\end{equation}
$\pwall$, $\pscint$ and $\kioq$ are calculated through Monte Carlo
simulations. $\kvol$ is determined through experimental data.

\subsubsection{Ionization quenching}\label{sec:quenching}

In this study, ionization quenching was included as a correction factor and was
calculated through Monte Carlo simulations following the formalism proposed by
Santurio et al \cite{Santurio.2020}. Using the empirical Birks law, which
express the light yield per unit path length as a function of the scintillator
efficiency and the quenching parameter ($kB$), the ionization quenching
correction factor is the light yield ratio with and without quenching:
\begin{equation}
    \kioq = \sum_{n=1}^N \left(
    \int_{E_{\min,n}}^{E_{\max,n}} \frac{1}{1+kB\cdot L_{\Delta}(E)}\dd{E}
    \right)
\end{equation}
where $N$ denotes the number of charged particles interacting within the PSD
sensitive volume, $E_{\mathrm{min},n}$ and $E_{\mathrm{max},n}$ indicate the
minimum and maximum energies of the \emph{nth} particle while within the
sensitive volume of the detector and $L_\Delta(E)$ is the restricted linear
electronic stopping power with an energy cutoff value $\Delta$.

The $kB$ value for the plastic scintillator was based on findings of Santurio et
al. \cite{Santurio.2020}, for a polystyrene-based scintillator. Here,   
$kB$ was set to be 0.019 \si{\centi\meter\per\mega\electronvolt} for a
conservative estimate of ionization quenching effects. In addition, simulations
with $kB = 0.01$ \si{\centi\meter\per\mega\electronvolt} were also made to asses
the sensitivity to this parameter on $\kioq$. 

\subsubsection{ Positional uncertainty ($\kpos$) and volume averaging ($\kvol$) corrections}
\label{sec:kvol}

Lechner et al. proposed an analytical formalism to evaluate the effects of
detector positional uncertainties on small field output factors
\cite{Lechner2020}. They used second-order polynomial fit to measured dose
profiles and positional uncertainty probability distribution functions to
determine the expectation value of the measured dose in small fields. The
approach of Lechner et al. is herein expanded to include the convolution arising
from volume averaging occuring within the sensitive volume of the detector. This
way, $k_\mathrm{pos}$ and $k_\mathrm{vol}$ correction factors can both be
determined using the same analytical approach.

Three hypotheses are made in order to determine $k_{pos}$ and $k_{vol}$. First,
the two-dimensional relative dose in a plane perpendicular to the beam's axis in
homogeneous water (i.e. in the absence of a detector) can be described by the
product of two one-dimensional second-order polynomial functions:
\begin{equation} \label{eq:dxy}
    D(x,y) = f(x) g(y) = (a_0 + a_1x +a_2x^2) (b_0 + b_1y +b_2y^2).
\end{equation}
This quadratic approximation is reasonable near the central axis of the beam and
has a maximum value at point $(x_\mathrm{max},y_\mathrm{max})$. Second, the
sensitive volume orientation and detector geometry are assumed to be known
precisely and the measured signal is assumed to be proportional to the integral
of the two-dimensional dose distribution weighted by the height of the detector
sensitive volume over a cross sectional area, $A$ of the detector in the $x-y$
plane:
\begin{equation} \label{eq:mxy}
    M(x_0,y_0) = \int_{A} D(x,y) h(x-x_0,y-y_0) dx dy,
\end{equation}
where $h(x-x_0,y-y_0)$ is the normalized height of the detector at position
$(x-x_0,y-y_0)$ from the detector's center, $(x_0,y_0)$. The normalization
serves to get a unit measurement signal for a uniform unit dose distribution.
Finally, the actual detector position is assumed to be known with an uncertainty
defined by a two-dimensional square probability density function of half-widths
$w_x$ and $w_y$. Therefore, the expectation value of a measurement signal at
($x_0,y_0$) can be determined by:
\begin{equation} \label{eq:expmxy}
    \left\langle M(x_0,y_0) \right\rangle = \frac{1}{4 w_x w_y} \int_{-w_x}^{w_x} \int_{-w_y}^{w_y} M(x+x_0,y+y_0) dx dy.
\end{equation}
Using these equations, the positional uncertainty and volume correction factors
are:
\begin{equation} \label{eq:kpos_def}
    k_{pos} = \frac{M(x_\mathrm{max},y_\mathrm{max})}{\left\langle M(x_\mathrm{max},y_\mathrm{max})\right\rangle},
\end{equation}
\begin{equation} \label{eq:kvol_def}
    k_{vol} = \frac{D(x_\mathrm{max},y_\mathrm{max})}{M(x_\mathrm{max},y_\mathrm{max})},
\end{equation}

The height function, $h$ and detector cross-sectional area, $A$ depend on the
detector sensitive volume geometry and orientation. Detector sensitive volume
geometries, as defined by the manufacturers, may be either spherical,
cylindrical or cylindrical with a half-spherical tip. Orientations are either
with the detector symmetry axis parallel or perpendicular to the beam axis (i.e.
parallel to the $z$ or $y$ axis respectively). Knowing the geometry and
orientation of a given detector, $M$ and $\left\langle M\right\rangle$ can be
determined analytically in terms of the $a_i$ and $b_i$ of eq.~(\ref{eq:dxy}) as
detailed in Supplementary material section~\ref{ap:detshape}. These analytical
equations combined with eqs.~(\ref{eq:kpos_def}) and (\ref{eq:kvol_def}) are
used to evaluate $k_\mathrm{pos}$ and $k_\mathrm{vol}$ correction factors.

Cross-line ($x$) and in-line ($y$) profiles were acquired with multiple
detectors and fitted simultaneously to equation~(\ref{eq:mxy}) to determine the
$a_i$ and $b_i$ parameters. Because the two dimensional dose function of
eq.~(\ref{eq:dxy}) is independent of detector geometry and orientations, the
fitted parameters obtained by fitting experimental data from one detector can be
used to evaluate correction factors for any other detector. Similarly to the
work by Lechner et al.~\cite{Lechner2020}, the variance of $M(x_0,y_0)$ can also
be analytically obtained from an integral equation similar to
equation~(\ref{eq:expmxy}) but for: $\left\langle M^2 \right\rangle -
\left\langle M\right\rangle^2$. This variance has been used to evaluate the
statistical distribution uncertainty on $k_\mathrm{pos}$.

\subsubsection{Monte Carlo simulations }

Monte Carlo calculations of absorbed dose-to-detector were performed with the
user code  \texttt{egs\_chamber} \cite{Wulff.2008} from EGSnrc
\cite{Kawrakow.2000, Kawrakow.20000d}. The phase space of the Varian Truebeam
6~MV photon beam above the jaws was provided by the manufacturer. To generate
phase spaces at specific field size, the jaws were simulated accordingly to the
manufacturer specifications with \texttt{BEAMnrc} \cite{rogers1995beam}. New
phase spaces at specific field sizes were scored at 75 cm from the source. These
phase spaces were validated by comparing simulated profiles and output factors
to published data \cite{Constantin.2011, Santurio.2020} and experimental
profiles. 

The PSD described in section~\ref{sec:hyperscint} was simulated based on data
provided by the manufacturer. The PSD was placed inside a water phantom (60
$\times$ 60 $\times$ 40~\si{\cubic\centi\meter}) at 10~\si{\centi\meter} depth.
The SSD was 90~cm, and field sizes were 30, 10, 8, 6, 5, 4, 2, 1, 0.8, and
0.6~cm. Monte Carlo parameters are presented in table \ref{tab:MC}, along with
the variance reduction techniques used to reduce calculation time. The number of
histories is such that the uncertainty of each Monte Carlo simulation is between
0.1-0.2\%.

\begin{table}[ht]
\caption{Record of Monte Carlo simulations characteristics and parameters used in EGSnrc.}
\resizebox{\linewidth}{!}{%
\centering 
\begin{tabular}{ll}
\hline\hline
Item & EGSnrc\\
\hline
MC code                      & \texttt{egs\_chamber}, EGSnrc 2021 release\\ 
Validation                   & Previously validated by Wulff et al \cite{Wulff.2008} \\
Scored quantities & Dose, light yield and cummulative dose\\
Source description & Phase space of a Varian Truebeam 6 MV photon beam\\
Cross sections      &  Default \\  
Transport parameters & Photon cutoff  0.001  keV \\ 
& Electron cutoff 0.512 keV \\
Variance reduction techniques & Photon cross-section enhancement\\
& Intermediate phase-space storage \\ 
& Range rejection and Russian Roulette\\
& ESAVE=ECUT=0.512 MeV\\ 
Statistical uncertainty & 0.1-0.2 \% in the sensitive volume \\ 
Statistical methods & Standard deviation of independent parallel simulations \\ 
\hline\hline
\end{tabular}%
}
\label{tab:MC}
\end{table}

For ionization quenching calculation, two Ausgab objects (AO) were implemented
to compute the light yield and the cumulative dose spectrum. These AO were
constructed in compliance with the guidelines presented in reference
\cite{Santurio.2020} and corroborated through direct communication with the
author.

%% file: content/method/meth_detectors.tex
\subsection{Detectors} \label{sec:detectors}

\subsubsection{Plastic scintillation detector system}\label{sec:hyperscint}

The Hyperscint RP-200 is a multi-channel scintillation dosimetry platform from
Medscint Inc. One channel was used in this study to read a PSD (Medscint's
\probe{}) which is composed of a proprietary plastic scintillator coupled to a
20~m clear plastic optical fiber guiding the optical signal outside the
treatment room to the spectral optical reader. The sensitive volume has a
cylindrical shape of 1~mm of diameter by 1~mm length. The detector calibration
was performed as described by the manufacturer. It includes measurements of the
individual spectral components from the scintillation, Cherenkov, fluorescence
and spectral attenuation of the optical fiber in order to allow for a correct
stem removal. To do so, the linac kV source was used to separate scintillation
and fluorescence spectra. Then, the MV beam was used to extract the Cherenkov
spectrum at two positions on the optical fiber to account for spectral
attenuation. Finally, the MV beam is also used to normalize the scintillation
signal which is proportional to dose. All PSD dose measurements were acquired at
a frame rate of 1 Hz. Seven probes, each with a 1 mm diameter and of the same
model, were used paired with either of the two available readers.

\subsubsection{IC and SS detectors}

Detector-specific field output correction factors have been extracted for
several detectors of different sizes and properties, with dimensions and
characteristics well-suited for small field dosimetry, as recommended in TRS-483
\cite{international2017iaea} report. The following micro-ionization chambers
were used: two IBA Razor Nano chambers (IBA RAZNC), two IBA Razor chambers (IBA
RAZC) and two Standard Imaging Exradin A26 (SI A26) chambers. Furthermore, two
synthetic micro diamond detectors from PTW, model 60019 (PTW 60019), and three
unshielded IBA Razor diodes (IBA RAZD) were also used. Relevant detector
physical properties are listed in table~\ref{tab:det}. Charges were collected
with a Supermax\textsuperscript{TM} electrometer (Standard Imaging, Middleton,
WI, USA) for all detectors except for the PSD that used a dedicated spectral
reader. Ionization chambers were operated at $\pm$300~V.

\begin{table}[ht]
\centering 
\caption{Physical properties of IC and SS detectors studied for detector-specific field output correction factor.}

\resizebox{\linewidth}{!}{%
\begin{tabular} {ccccccc}
\hline\hline
Name & Type  & Active volume & Active volume  & Sensitive  & Wall material  & Central electrode \\
 &   &  & length/radius (mm) & material (g/cm$^3$) & (g/cm$^3$)  & material (g/cm$^3$) \\
\hline
IBA RAZNC & Ion chamber  & spherical, 3 mm$^3$ & - / 1.0 & air (0.001) & C552 (1.76)  & graphite (2.26 ) \\ 
IBA RAZC & Ion chamber  & cyl/half-sph, 11 mm$^3$ & 2.6 / 1.0 & air (0.001) & C552 (1.76)  & graphite (2.26) \\ 
SI A26 & Ion chamber  & spherical, 15 mm$^3$ &  - / 1.65 & air (0.001) & C552 (1.76 ) & C552 (1.76) \\ 
IBA RAZD & Unshielded diode & cylindrical, 0.006 mm$^3$ &  0.02 / 0.3 & p-type silicon (2.33) & ABS (1.05) & - \\ 
PTW 60019 & Synthetic diamond  & cylindrical, 0.004 mm$^3$ & 0.001 / 1.1 & diamond (3.52) & RW3 (1.05)  & - \\ 
\probe{} & PSD  & cylindrical, 0.79 mm$^3$ & 1 / 0.5 & Polystyrene with dopants (1.06) & Nylon (1.01)  & - \\ 
\hline\hline
\end{tabular}%
}
\label{tab:det}
\end{table}

For ICs, measurements at both +300~V and -300~V bias were performed for each
field size in order to determine $\kpol$. The ion recombination correction was
applied as described in section~\ref{sec:meas_corr} to all detectors using a $B$
parameter determined by fitting $k_\mathrm{ion}$ values obtained from the Boag
two-voltage (150~V/300~V) method of a Standard Imaging A26 IC.

%% file: content/method/meth_measurements.tex
\subsection{Measurement methodologies}

\subsubsection{Experimental setup}

All measurements were performed on a Varian Truebeam linac (Varian Medical
Systems, Palo Alto, CA, USA) using a 6~MV flattened photon beam. Output factor
measurements were performed by placing the effective point of measurement of a
given detector at the linac isocenter, at 10~cm water depth in a motorized IBA
Smartscan water tank (IBA Dosimetry, Schwarzenbruck, Germany). The SSD was set
to 90~cm and was adjusted using the magnetic front pointer with the gantry
placed upright at 0\textdegree. Each detector's symmetry axis was aligned
parallel to the beam axis, with the stem at deeper depths for all detectors
except PSDs. For PSD, the optical fiber was aligned perpendicular to the beam
axis (in-line direction), in order to minimize stem signal. 

Centering of detectors was performed with the ``11-points'' methodology,
described in section \ref{sec:centering_method}. Measurements were performed for
twelve jaw-delimited square fields with side lengths of 0.6, 0.8, 1.0, 1.5, 2.0,
3.0, 4.0, 5.0, 6.0, 8.0, 10.0 and 30.0~cm. One hundred Monitor units (MU) were
delivered for each irradiation and at least 3 irradiations were performed for
each field size. The 10~$\times$~10~cm$^2$ field size was considered the
reference field size ($\mathrm{msr} = \mathrm{ref} = 10 \times 10$~cm$^2$) and
has been measured frequently in order to be used as a normalization field and to
reduce machine output, temperature and pressure drifts.

\subsubsection{Equivalent square field size and detector centering method} 
\label{sec:centering_method}

For each nominal field size, actual measured field sizes, $FS_x$ and $FS_y$,
were converted to the equivalent measured square small field size $\SCLIN$
following the approach adopted by TRS-483~\cite{international2017iaea} where
$\SCLIN$ is given by: 
\begin{equation} \label{eq:sclin}
    S_{\text{clin}}=\sqrt{FS_x \times FS_y},
\end{equation}
where $FS_x$ corresponds to the radiation field full width at half maximum
(FWHM) in cross-line direction, $x$, and $FS_y$ (FWHM) for in-line direction,
$y$, perpendicular to the former. $FS_x$ and $FS_y$ were determined for all
field sizes using an 11-points technique allowing both equivalent field size
measurement and precise detector centering. First, the detector is visually
centered. Then, for cross-line orientation, in a step-and-shoot operation, two
measurements of 100~MU are acquired on each side of the field at positions where
the dose is estimated to be close to 30\% and 70\% of the profile's maximum
dose. Assuming linear profile slopes in these penumbra regions, the cross-line
field center position is determined. The detector is moved at that position and
then a fifth measurement is acquired for cross-line profile values normalization
and FWHM ($FS_x$) calculation. The procedure is repeated for in-line
orientation, adding five other measurements. Finally, the 11$^{th}$ measurement
is acquired at the precise field center which is now known in both directions.

\subsubsection{Analytical functions of $\Mratio$ and $\kfclin$}

In order to be able to calculate field output correction factor for any given
$\SCLIN$, values from evaluated $\Mratio$ and $\kfclin$ at measured $\SCLIN$
were fitted with analytical functions optimized with a bounded non-linear least
square trust region reflective algorithm~\cite{SciPy2020, Branch1999}. First,
$\Mratio$ is fitted with the function given in Sauer and
Wilbert~\cite{Sauer.2007}: 

\begin{equation} \label{eq:fit_OF}
\Mratio\left(S_{\text {clin }}\right)=P_{\infty} \frac{S_{\text {clin }}^n}{l^n+S_{c l i n}^n}+S_{\infty}\left(1-e^{-b \cdot S_{\mathrm{clin}}}\right),
\end{equation}
where $l$, $n$, $b$, $P_{\infty}$, $S_{\infty}$ are fitting parameters.
Eq.~(\ref{eq:fit_OF}) is normalized so that $\Mratio(10~cm)$ is unity.
$\kfclin$ is then fitted with the function given in TRS-483
report~\cite{international2017iaea}: 
\begin{equation} \label{eq:fit_kfclin}
    \kfclin\left(S_{\mathrm{clin}}\right)=\frac{1+a_4  e^{-\frac{10-a_1}{a_2}}}{1+a_4  e^{-\frac{S_{\mathrm{clin}}-a_1}{a_2}}}+a_3 \left(S_{\mathrm{clin}}-10\right),
\end{equation}
where $a_1$, $a_2$, $a_3$ and $a_4$ are fitting parameters.

%% file: content/method/meth_uncertainties.tex
\subsection{ Uncertainties}

A careful uncertainty analysis is necessary to quantitatively evaluate the   
field output correction factors obtained for Medscint's \probe{} and compare
them to other detectors. Uncertainty analysis was performed for (1) the
reference PSD $\kfclin$, (2) the $\kfclin$ derived with other detectors and (3)
the measured field output factor. 

\subsubsection{\probe{} uncertainty on $\kfclin$}\label{sec:unc_psd}

Field output correction factors for Medscint PSD detector are determined with
eq.~(\ref{eq:kfclin_psd}). Three parameters come from Monte Carlo simulations:
$P_{wall}$, $P_{scint}$ and $k_{ioq}$ ratios. Type~A statistical uncertainties
(\emph{i.e.}  uncertainties derived using statistical analysis of a series of
observations) on these three ratios are approximately $0.1\%$, $0.1\%$ and
$0.35\%$ respectively. Two sources of type~B uncertainties were also considered
on these ratios. The first arises from inaccuracies on physical data used in
cross-sections. Studies have investigated the impact of systematic uncertainty
from physical data on the beam quality correction factor $k_{Q,Q_0}$ for
ionization chambers \cite{Muir.2010, Wulff.2010}. Wulff et al.~\cite{Wulff.2010}
suggested a systematic uncertainty of 0.2\% for 6~MV photon beams. However, this
value is dominated by the uncertainty on graphite $I$-value. Despite graphite
being absent from PSDs, this systematic uncertainty was used in this study.
Furthermore, changes in beam quality resulting from changes in field sizes (i.e.
changes between $Q_{msr}$ and $Q_{clin}$) are expected to be smaller than
changes between $Q_0$ and $Q$ that were the focus of the work of Wulff et al.
Thus, this 0.2\% type~B uncertainty for cross sections inaccuracies is a
conservative estimate. The second source of type~B uncertainty arises from
geometric and density differences between the actual detector and the Monte
Carlo model. This uncertainty was determined, as suggested by others
\cite{Francescon.2014, Muir.2010}, by performing simulations with sensitive
volume with diameters varying by $\pm5\%$. 

The last factor affecting $\kfclin$ uncertainty is the experimental
$\left(k_\mathrm{vol}\right)_{Q_\mathrm{clin}, Q_\mathrm{msr}}^{f_\mathrm{clin},
f_\mathrm{msr}}$ ratio. Uncertainties on this factor come from two sources:
(1)~type A statistical component evaluated with the standard deviation of the
mean of equation~(\ref{eq:kvol_def}) from several detectors profile
measurements, and (2)~type B component that is due to geometric uncertainties,
and is evaluated, as previously, with a $\pm5\%$ geometric parameters variation.
The total combined uncertainty on reference PSD field output correction factor
is evaluated by summing in quadrature all those contributions.

\subsubsection{Other detectors uncertainty on derived $\kfclin$}\label{sec:unc_det}

Any other detector field output correction factor derived from comparative
measurements with Medscint's \probe{}, using equation~(\ref{eq:kfclin_det2}) depends
on five factors for which the uncertainties must be evaluated. The first factor
is the PSD's $\kfclin$ (see section~\ref{sec:unc_psd}). The four other factors
are for the corrected measurements. The relative uncertainty of each parameters
required for measurement correction (i.e. eq.~(\ref{eq:mcorr_ic}) for IC,
eq.~(\ref{eq:mcorr_ss}) for SS and eq.~(\ref{eq:mcorr_psd}) for PSD) expressed
as a ratio between $\mathrm{clin}$ and $\mathrm{msr}$ fields must be considered. 

Type~A statistical uncertainty of $M_{raw}$ ratios (or $\kpol M_{raw}$ ratios
for IC detectors) was taken as the standard deviation of the mean from multiple
measurements taken with several detectors of the same model. Those variations in
measured raw values are expected to come mainly from readout and linac dose
short term reproducibility as well as detector and jaws positioning
reproducibility. Combined type B uncertainty for temperature, pressure, humidity,
readout, background and machine drift correction ratios have been estimated to
be at most $0.1\%$. Type B uncertainty on $k_\mathrm{ion}$ correction ratio has
been estimated using a worst case scenario from equation~(\ref{eq:kion_ratio2})
with $B$ ranging from lowest to highest estimated values between all tested
SSDs. Type~A and~B uncertainties in $k_\mathrm{pos}$ ratio were evaluated
similarly as those on $k_{vol}$ ratio described in section~\ref{sec:unc_psd}. 

For $k_\mathrm{stem}$ of PSD measurements, a sensitivity analysis estimated the
overall uncertainty. Using the experimental noise of each wavelength of each
spectrum acquired during the calibration procedure, two hundred different
calibration sets were generated. Each set was applied to correct the stem for
each field size measurement. The relative standard deviation of the distribution
at larger field sizes was found to be lower than 0.1\%. For other types of
detector, the $k_\mathrm{stem}$ uncertainty is estimated to be lower than 0.1\%.
Furthermore, vertical orientation were used to minimize $k_\mathrm{stem}$
deviation between field sizes. Conservatively, a value of 0.1\% was used.

Finally, because the derived field output correction factors are associated,
tabulated and fitted with an effective square field size dependence, the
uncertainty on the field size determination must be propagated to the $\kfclin$
uncertainty. Therefore, the derivative of the fitted corrected measurement ratio
with field size is calculated and multiplied by the field size determination
uncertainty to obtain that component of $\kfclin$ total uncertainty. The
equivalent square field size determination uncertainty have been evaluated by
taking the standard deviation of a series of five consecutive $FS_x$ and $FS_y$
measurements using the method described in section~\ref{sec:centering_method}
for a fixed 0.6$\times$0.6~cm$^2$ nominal field. The effective square field size
uncertainty is obtained by combining uncertainties on $FS_x$ and $FS_y$ using
the derivative of equation~(\ref{eq:sclin}).

Estimation of the detector central position uncertainty is done by using the
standard deviation of the calculated central position from the previously
described series of five consecutive field size measurements with the same fixed
nominal field. This position uncertainty must further be combined with the
inherent positioning system uncertainty that will ultimately affect the final
detector effective position. The total positioning uncertainty is assumed here
to have a square probability distribution with half-width $w_x$ and $w_y$. 

The uncertainty on corrected measurement ratio due to detector positioning is
estimated with the standard deviation of the mean of $k_\mathrm{pos}$ (evaluated
from $w_x$ and $w_y$ values) for the number of times a detector has been
positioned for that field. To estimate the uncertainty on corrected measurement
ratio due to jaws positioning when a detector is centered after each jaw
movement, the field size uncertainty due to jaws repositioning is first
evaluated then multiplied by the derivative of the corrected measurement ratio
at the measured field size. The field size uncertainty due to jaw repositioning
is evaluated with a series of five consecutive field size measurements with the
same 0.6$\times$0.6~cm$^2$ nominal field but with jaw movements between each
field size measurement.

\subsubsection{Field output factor uncertainty}\label{sec:OF_MC_uncert}

The last part of the uncertainty analysis consists of evaluating the measured
field output factor uncertainty. Following equation~(\ref{eq:OF_TRS}), three
terms are involved. The first one is the uncertainty on the detector $\kfclin$
ratio (see sections~\ref{sec:unc_psd} and~\ref{sec:unc_det}). The two other
factors are the uncertainties on $\mathrm{clin}$ and $\mathrm{msr}$ corrected
measurements. The combined uncertainty of the corrected measurement ratio is
determined as described in previous sections. The total combined uncertainty on
field output factor measurements is evaluated by summing in quadrature the three
terms contribution.

In the case where $\kfclin$ correction factor is extracted from the fitted
function in equation~(\ref{eq:fit_kfclin}), the uncertainty on the extracted
value can be evaluated with a Monte Carlo statistical sampling method. The curve
fitting procedure was repeated a thousand times with $\kfclin$ values sampled
from a gaussian distributions with measured $\kfclin$ as mean with a standard
deviation equal to the uncertainty calculated from one of the last two sections.
The estimated uncertainty is divided in two parts; the first involves all
components of random nature potentially affecting differently each
$\kfclin(S_\mathrm{clin})$ and the second involves the remaining uncertainty
components that could affect all $\kfclin(S_\mathrm{clin})$ values
systematically. Therefore, the resampling of $\kfclin$ values at each iteration
is performed in two steps. A first random number is used to sample the relative
deviation affecting values at all $S_\mathrm{clin}$, then new random numbers are
drawn for each $S_\mathrm{clin}$ and are used to sample the random portion of
the estimated uncertainties. The two values are added to obtain the resampled
$\kfclin(S_\mathrm{clin})$ values for each iteration. Standard deviation of the
resulting fits at every  
0.5~mm for $S_\mathrm{clin}$ between 0.5~cm and 40.0~cm were determined and then
used as $\kfclin$ fit uncertainties.

%% file: content/method/meth_kfclin_all.tex
\subsection{$\kfclin$ determination for all detectors}

This section describes the process used to determine $\kfclin$ correction
factors. This is first done for the detector used as a reference (Medscint
\probe{}) as detailed in section~\ref{sec:kfclin_psd} and then for all other
detectors.  

Following this and using eq.~(\ref{eq:kfclin_psd}), $\kfclin$ values and the
associated uncertainties of the reference PSD can be determined for the selected
set of field sizes. These values are then fitted to a second degree polynomial
to interpolate them at any field size. Once $\kfclin$ is known for the
reference, the other detectors, readout equipment and positioning system are
characterized. For each detector (including the reference PSD), output factor
measurements are acquired for all field sizes. Each measurement is preceded by
field size and field center position determination (see
section~\ref{sec:centering_method}). From these, field output correction factors
and their uncertainties can be determined using equation~(\ref{eq:kfclin_det2}).
Finally, $\kfclin$ data is fitted using eq.~(\ref{eq:fit_kfclin}) and the
associated fit uncertainties can be determined with the sampling method
described in section~\ref{sec:OF_MC_uncert}. Comparison with literature data can
then be performed at specific field sizes, using the fit data and two-sided
unpaired Welch's unequal variances t-test \cite{Welch1947,SciPy2020}.

%% file: content/results.tex
\section{Results}

\input{content/results/res_meas_corrections.tex}

\input{content/results/res_kfclin_psd.tex}

\input{content/results/res_uncertainties.tex}

\input{content/results/res_kfclin_dets.tex}

%% file: content/results/res_meas_corrections.tex
\subsection{Measurement corrections}

\begin{figure}
    \centering
    \includegraphics[width=0.7\textwidth]{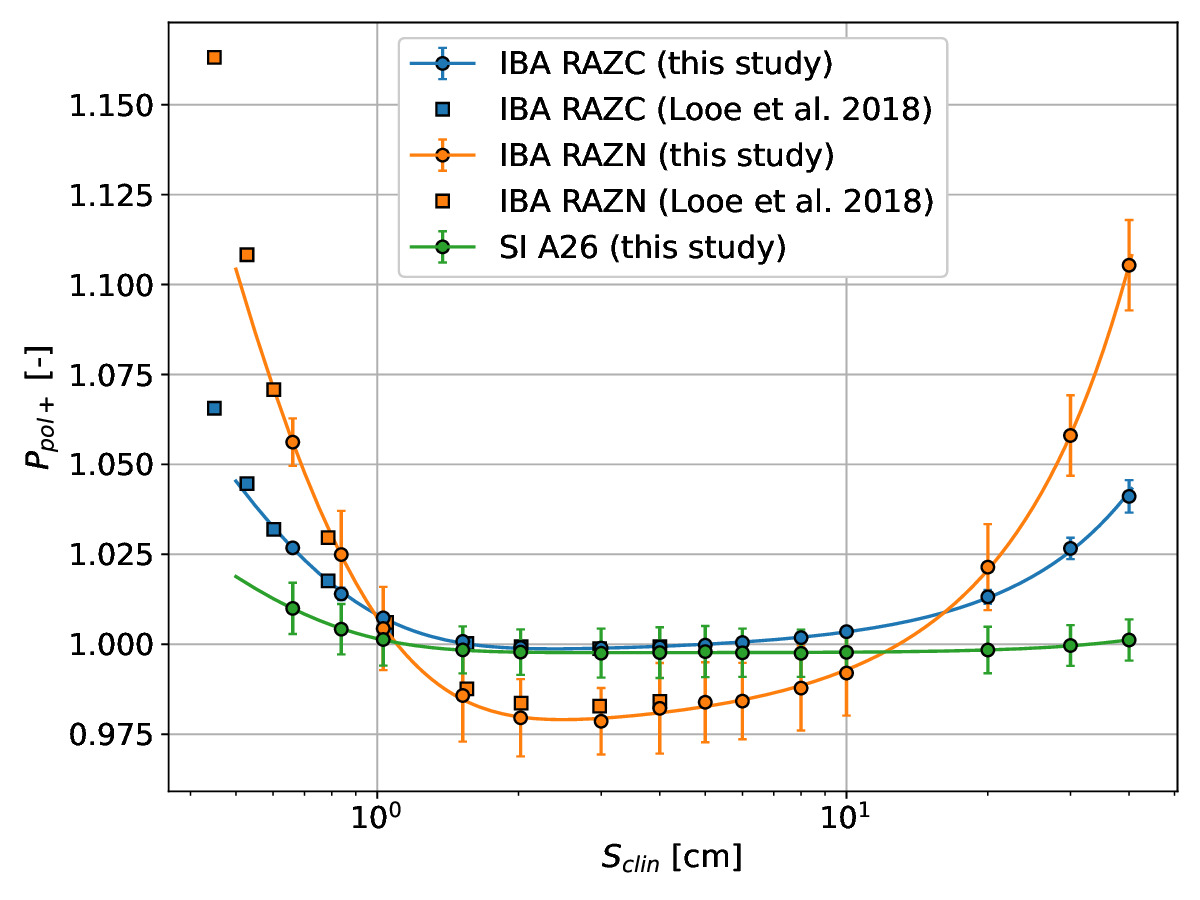}
    \caption{Comparison of polarity correction factor, $\kpol$, for the
    ionization chambers: IBA Razor,  IBA Razor Nano and Standard Imaging Exradin
    A26. The fit models $\kpol$ behavior with a modified exponential base,
    complemented by linear and quadratic terms. Results extracted from Looe et
    al.~\cite{Looe2018} are also added to the figure but reversed according to
    the different polarity sign convention used.}
    \label{fig:ppol_per_detectors}
\end{figure}

The polarity correction factor, specifically for measurements with positive
polarity in the vertical orientation, vary with field size. This relationship is
depicted in figure \ref{fig:ppol_per_detectors} for three detectors: IBA Razor
Chamber,  IBA Razor Nano Chamber, and Standard Imaging Exradin A26. Results for
IBA chambers are in good agreement with those from Looe et al.~\cite{Looe2018}. 
Among studied detectors, the IBA Razor Nano's $\kpol$ exhibits the largest
variation with field size with values increasing beyond 10\% at both small and
large fields. The SI Exradin A26 had the $\kpol$ values closer to unity with
deviations mainly observed at the smallest fields.

Table~\ref{tab:B} presents the fitted parameter $B$ from eq.~(\ref{eq:kion_ratio2}) as
well as calculated $\kion$ ratios of three field sizes for IC and SS detectors.

\begin{table}
\caption{Values for parameter $B$ determined using eq.~(\ref{eq:kion_ratio2}) and
$(k_\mathrm{ion})_{Q_\mathrm{clin},Q_\mathrm{msr}}^{f_\mathrm{clin},f_\mathrm{msr}}$ for each detector at $0.6, 1.0$ and $2.0$~cm side of squared field sizes. Uncertainties are shown in brackets and represent
absolute uncertainties in the last digit.}
    \centering
    \begin{tabular}{ccccc}
    \hline\hline
          &  & & $(k_\mathrm{ion})_{Q_\mathrm{clin},Q_\mathrm{msr}}^{f_\mathrm{clin},f_\mathrm{msr}}$ & \\
        Detector & $B$ & (0.6 $\times$ 0.6~\si{\centi\meter\squared}) & (1 $\times$ 1~\si{\centi\meter\squared}) & (2 $\times$ 2~\si{\centi\meter\squared})\\
    \hline   
        IBA RAZNC & 0.0035 (7) & 0.9984 (3) & 0.9990 (2) & 0.9993 (1) \\
        IBA RAZC & -0.001 (1) & 1.0004 (4) & 1.0003 (3) & 1.0002 (2) \\
        SI A26 & 0.0004 (1) & 0.9998 (1) & 0.9999 (1) & 0.9999 (1) \\
        IBA RAZD & -0.005 (1) & 1.0022 (1) & 1.0016 (1) & 1.0011 (1) \\
        PTW 60019 & -0.003 (2) & 1.0011 (5) & 1.0008 (3) & 1.0005 (2) \\
    \hline\hline
    \end{tabular}
    \label{tab:B}
\end{table}


Calculations for $\kpos$ for all detectors geometries and orientations were
found to be 1.000 with negligible uncertainty. Furthermore, for Medscint's
\probe{}, the sensitivity study testing the impact of $\pm$5\% variation in the
sensitive volume radius on $\kpos$ did not reveal any change in $\kpos$.
Therefore, a $\kpos$ value of 1.000 was used throughout this work.

\subsubsection{Output factors corrected for type I influence factors}

\begin{figure}[ht]
\centering
\includegraphics[width=0.8\textwidth]{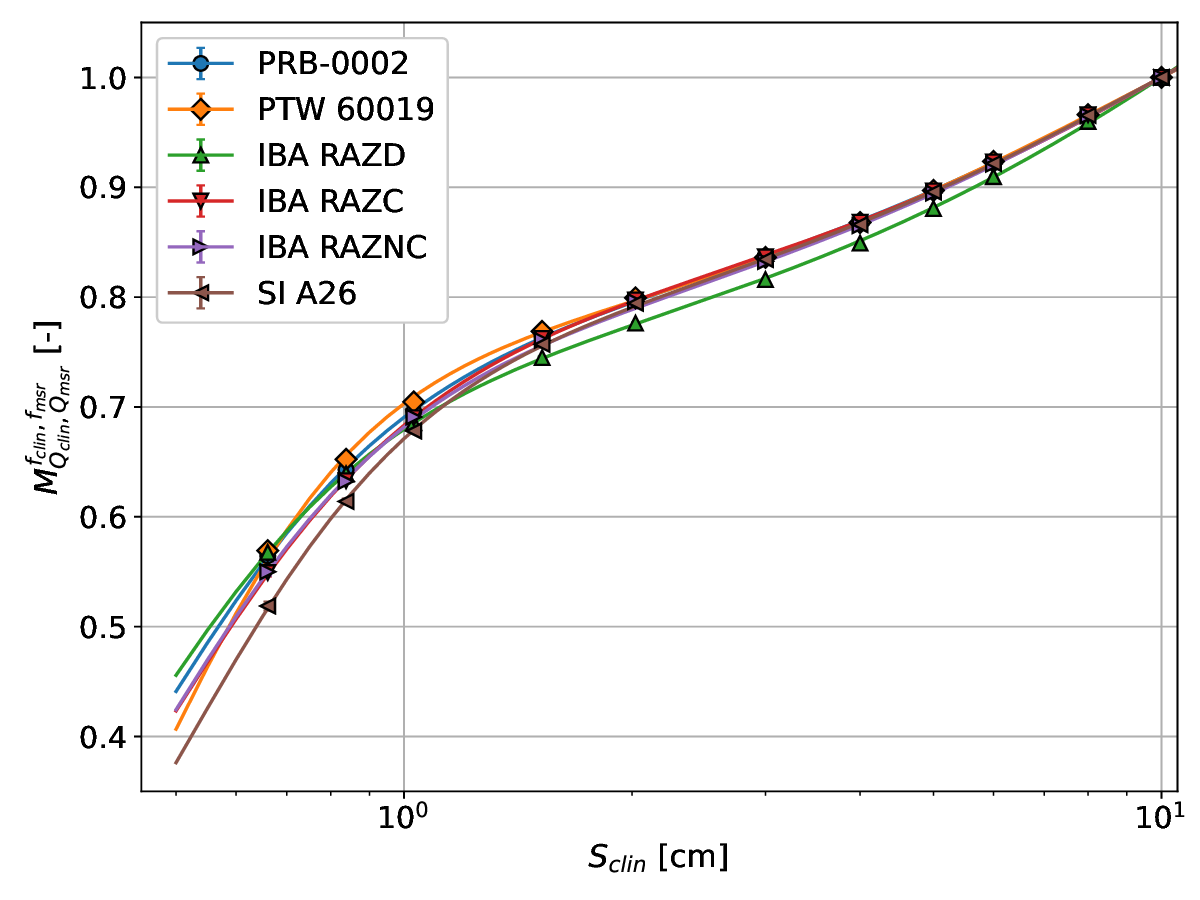}

\caption{Measurement ratios corrected only for type I factors, $\Mratio$. The
regression combines a sigmoid saturation model and an exponential decay
component.}
\label{fig:OF_per_detectors}
\end{figure}

Measurements ratio for all detectors corrected by influence factors that are not
affected by change in beam quality (i.e. type I factors) as detailed in
eqs.~(\ref{eq:mcorr_ic})-(\ref{eq:mcorr_psd}), $\Mratio$, is presented in
Fig.~\ref{fig:OF_per_detectors}. The fit was performed using
eq.~(\ref{eq:fit_OF}). The corrected measurement ratios of the IBA Razor diode
distinctly deviates from the trend of the other detectors. This is possibly due
to an over-response to lower energy components present for fields larger than
5~$\times$~5~\si{\centi\meter\squared}. As field size decreases, there is a
pronounced dispersion in the $\Mratio$ values across detectors, with the SI A26
exhibiting the most significant variation.

%% file: content/results/res_kfclin_psd.tex
\subsection{Determination of $\kfclin$ for \probe{}}

\begin{figure}
    \centering
    \includegraphics[width=0.8\textwidth]{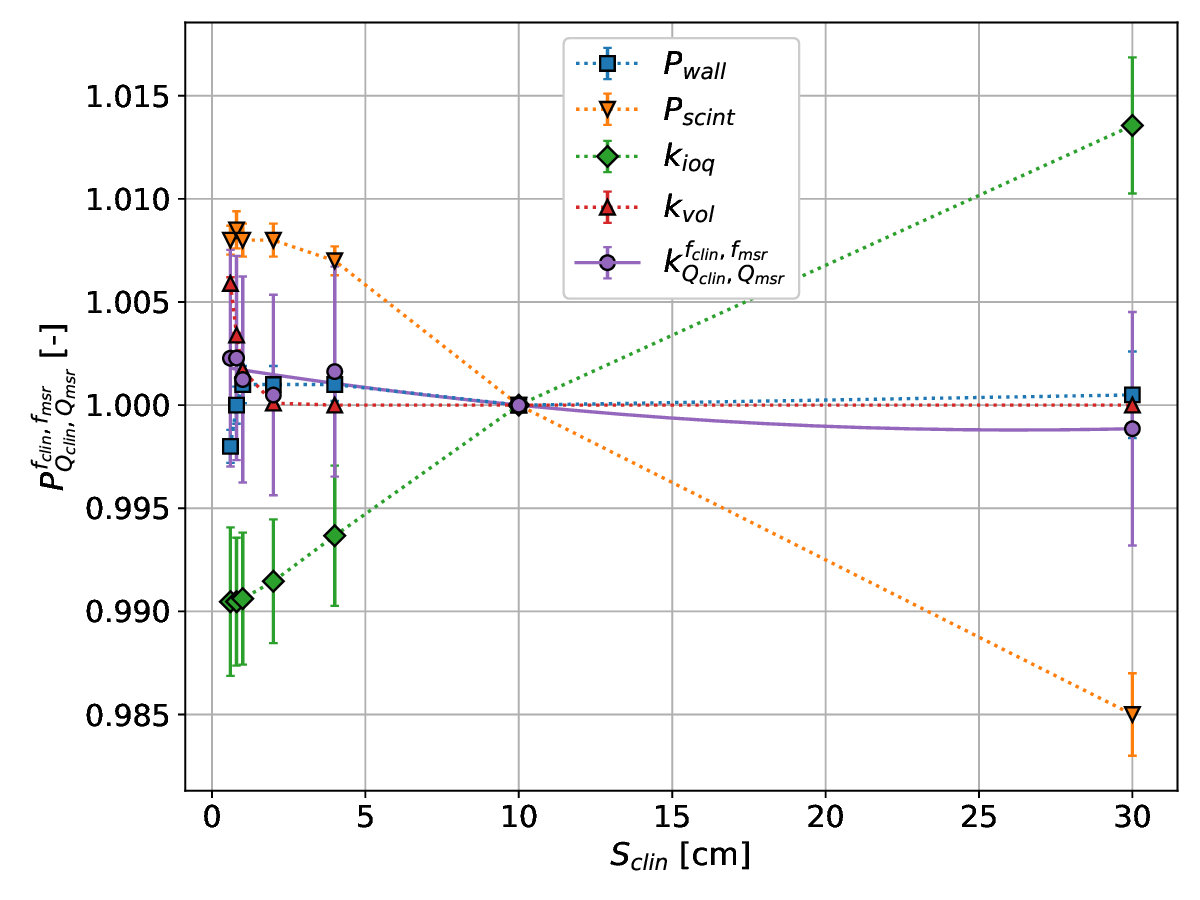}

    \caption{Monte Carlo calculated perturbation and correction factors of
    Medscint's \probe{}, including $\kfclin$. $\pwall$ accounts for the
    perturbations coming from the extracameral components, $\pscint$ for the
    perturbations due to variation in density and atomic composition. $\kioq$ is
    the ionization quenching correction factor and $\kvol$ is the volume
    averaging correction factor determined by dose profiles measurements. The
    machine specific reference, $\mathrm{msr}$, field is
    10$\times$10~\si{\centi\meter\squared}}
    \label{fig:perturbations}
\end{figure}

Medscint's \probe{} output correction factors, $\kfclin$ obtained from
eq.~(\ref{eq:kfclin_psd}), along with the correction and perturbation factors
($\kioq$, $\kvol$, $\pwall$, $\pscint$), are shown in
figure~\ref{fig:perturbations} for field  sizes ranging from 0.6~$\times$~0.6 to
30~$\times$~30~\si{\centi\meter\squared}. It can be seen that $\pscint$ and
$\kioq$ offset each other, resulting in \probe{}'s $\kfclin$ being primarily
influenced by $\kvol$. The uncertainties for $\pwall$ and $\pscint$ are both
0.1\% and uncertainty for $\kioq$ is 0.36\%, leading to a total uncertainty of
0.4\% on $\kfclin$. The sensitivity of $\kioq$ to variations in $kB$ (see
section~\ref{sec:quenching}) showed no statistical difference. 

$\kvol$ is derived using eq.~(\ref{eq:kvol_def}) in conjunction with measured
dose profiles. These profiles were measured with all detectors (see
Supplementary material section~\ref{ap:kvol_profiles}) and the value of $\kvol$
for \probe{} was taken as the average.

%% file: content/results/res_uncertainties.tex
\subsection{Uncertainties}

Uncertainty budget for $\kfclin$ and $\OF$ measurements with the \probe{} is
detailed in tables \ref{tab:unc_budget_kfclin_HS} and
\ref{tab:unc_budget_OF_HS}. Equivalently, uncertainty budget of the PTW 60019
microDiamond detector is shown in tables \ref{tab:unc_budget_kfclin_mD} and
\ref{tab:unc_budget_OF_mD} as another example. These budgets enumerate the
sources of uncertainty that contribute to the combined uncertainty, presented
with a one-sigma probability coverage (k=1), for three selected field sizes.

\begin{table}
\centering

\caption{Example uncertainty budget for determining $\kfclin$ with the PSD
\probe{}. This includes statistical uncertainties for each factor from Monte
Carlo simulations, overall uncertainty due to physical data (e.g.,
cross-section), and uncertainties between actual and modelled density and
geometry. For the hybrid method of calculating $\kvol$, uncertainties include
statistical variation from measurements and geometrical differences between
actual and theoretical geometries.}

\begin{tabular}{>{\arraybackslash}m{0.4\textwidth}>
{\centering}p{0.06\textwidth}>
{\centering}p{0.15\textwidth}>
{\centering}p{0.15\textwidth}>
{\centering\arraybackslash}p{0.15\textwidth}}
\hline\hline
\multicolumn{1}{c}{\multirow{2}{*}{Item}} & \multirow{2}{*}{Type} & \multicolumn{3}{c}{Field size {[}cm²{]}} \\
\multicolumn{1}{c}{} & & 0.6 $\times$ 0.6 & 1 $\times$ 1 & 2 $\times$ 2 \\ \hline
Monte Carlo & & & & \\ 
\hspace{8 pt}Statistical $P_\mathrm{wall}$ & A & 0.08\% & 0.09\% & 0.09\% \\ 
\hspace{8 pt}Statistical $P_\mathrm{scint}$ & A & 0.07\% & 0.08\% & 0.08\% \\ 
\hspace{8 pt}Statistical $k_\mathrm{ioq}$ & A & 0.36\% & 0.32\% & 0.30\% \\ 
\hspace{8 pt}Cross-section \cite{Wulff.2010} & B & 0.2\% & 0.2\% & 0.2\% \\ 
\hspace{8 pt}Geometry \& Density \cite{Francescon.2014} & B & 0.3\% & 0.3\% & 0.3\% \\ \hline
Experimental/Analytical & & & & \\ 
\hspace{8 pt}Statistical $k_\mathrm{vol}$ (7 dets) & A & 0.026\% & 0.009\% & 0.002\% \\
\hspace{8 pt}Geometry $k_\mathrm{vol}$ & B & 0.03\% & 0.005\% & 0.001\% \\
\hline
Combined uncertainty (k=1) & & 0.52\% & 0.50\% & 0.48\% \\ 
\hline
\hline
\end{tabular}
\label{tab:unc_budget_kfclin_HS}
\end{table}

\begin{table}
\centering

\caption{Uncertainty budget example for $\OF$ measurements with the PSD
\probe{}. Sources of uncertainty of $M_{f_\mathrm{clin},f_\mathrm{msr}}$,
$(M_\mathrm{raw})_{f_\mathrm{clin},f_\mathrm{msr}}$ for a single detector and
field determination are explicitly detailed. The $\star$ symbol denotes
uncertainties from seven detectors with two measurements each, and the $\square$
symbol indicates uncertainty from a single measurement with one detector. }

{\footnotesize
\begin{tabular}
{>{\arraybackslash}m{0.35\textwidth}>
{\centering}p{0.055\textwidth}>
{\centering}p{0.19\textwidth}>
{\centering}p{0.19\textwidth}>
{\centering\arraybackslash}p{0.17\textwidth}}
\hline\hline
\multirow{2}{*}{Item} & \multirow{2}{*}{Type} & \multicolumn{3}{c}{Field size {[}cm²{]}} \\
 &  & 0.6 $\times$ 0.6 & 1 $\times$  1 & 2 $\times$  2 \\
 \hline
$\kfclin$ PSD (see table \ref{tab:unc_budget_kfclin_HS}) $\square$ $\star$ & B & 0.52\% & 0.50\% & 0.48\% \\
\hline
$M_{f_{clin},f_{msr}}$ & & & & \\ 
Statistical $M_{raw}$ (1 meas) $\square$ & A & 0.29\% & 0.16\% & 0.11\% \\
Statistical $M_{raw}$ (7 dets $\times$ 2 meas) $\star$ & A & 0.08\% & 0.04\% & 0.03\% \\ 
$k_{stem}$ $\square$ $\star$ & B & 0.1\% & 0.1\% & 0.1\% \\ 
$k_{other}$ $\square$ $\star$ & B & 0.1\% & 0.1\% & 0.1\% \\ 
Statistical $k_{pos}$ (7 dets) $\square$ $\star$ & A & - & - & - \\
Geometry $k_{pos}$ $\square$ $\star$ & B & - & - & - \\
\hline
$(M_{raw})_{f_{clin},f_{msr}}$ (single det.) & & & & \\
Positioning system/FS center $w_x$/$w_y$ & B & & 0.1/0.1~mm & \\
Positioning system/FS center & B & 0.02\% & 0.01\% & - \\
Readout/Machine reproducibility & A & 0.1\% & 0.1\% & 0.1\% \\
Jaws reproducibility $FS_x$/$FS_y$ & A & & 0.02/0.05~mm & \\
Jaws reproducibility $S_\mathrm{clin}$ & A & 0.026~mm~(0.28\%) & 0.027~mm~(0.08\%) & 0.027~mm~(0.02\%) \\
Combined uncertainty (k=1) & & 0.30\% & 0.13\% & 0.10\% \\
\hline
Field size determination & & & & \\
Statistical $FS_x$/$FS_y$ (1 centering) & A & & 0.06/0.07~mm & \\
Statistical $FS_x$/$FS_y$ (7 centerings) & A & & 0.02/0.03~mm & \\
Inter-det-type $FS_x$/$FS_y$ & B & & 0.07/0.08~mm & \\
$S_\mathrm{clin}$ (1 centering) $\square$ & A & 0.071~mm~(0.75\%) & 0.071~mm~(0.22\%) & 0.071~mm~(0.05\%) \\
$S_\mathrm{clin}$ (7 centerings) $\star$& A & 0.053~mm~(0.57\%) & 0.053~mm~(0.16\%) & 0.053~mm~(0.04\%) \\
\hline
Combined uncertainty (k=1) $\square$ & &  0.97\% & 0.59\% & 0.51\% \\ 
Combined uncertainty (k=1) $\star$ & &  0.79\% & 0.55\% & 0.50\% \\ 
\hline
\hline
\end{tabular}
\label{tab:unc_budget_OF_HS}
}
\end{table}

\begin{table}
\centering

\caption{Uncertainty budget example for $\kfclin$ determination of the PTW60019
detector. Sources of uncertainty of $M_{f_\mathrm{clin},f_\mathrm{msr}}$,
$(M_\mathrm{raw})_{f_\mathrm{clin},f_\mathrm{msr}}$ for a single detector  and
field determination are explicitly detailed. The $\star$  denotes uncertainties
from two detectors with four measurements each, and the $\square$ indicates
uncertainty from a single measurement with one detector. }

{\footnotesize
\begin{tabular}{>{\arraybackslash}m{0.345\textwidth}>
{\centering}p{0.055\textwidth}>
{\centering}p{0.19\textwidth}>
{\centering}p{0.19\textwidth}>
{\centering\arraybackslash}p{0.175\textwidth}}
\hline\hline
\multicolumn{1}{c}{\multirow{2}{*}{Item}} & \multirow{2}{*}{Type} & \multicolumn{3}{c}{Field size {[}cm²{]}} \\
\multicolumn{1}{c}{} & & 0.6 $\times$ 0.6 & 1 $\times$ 1 & 2 $\times$ 2 \\
\hline
$\OF$ PSD (see table \ref{tab:unc_budget_OF_HS}) $\square$ $\star$ & B & 0.79\% & 0.55\% & 0.50\% \\
\hline
$M_{f_\mathrm{clin},f_\mathrm{msr}}$ PTW 60019 & & & & \\ 
Statistical $M_\mathrm{raw}$ (1 meas) $\square$ & A & 1.3\% & 0.42\% & 0.10\% \\ 
Statistical $M_\mathrm{raw}$ (2 dets $\times$ 4 meas) $\star$ & A & 0.46\% & 0.15\% & 0.04\% \\ 
$k_\mathrm{ion}$ $\square$ $\star$ & B & -\% & -\% & -\% \\ 
$k_\mathrm{stem}$ $\square$ $\star$ & B & 0.1\% & 0.1\% & 0.1\% \\ 
$k_\mathrm{other}$ $\square$ $\star$ & B & 0.1\% & 0.1\% & 0.1\% \\ 
Statistical $k_\mathrm{pos}$ (2 dets) $\square$ $\star$ & A & - & - & - \\ 
Geometry $k_\mathrm{pos}$ $\square$ $\star$ & B & - & - & - \\
\hline
$(M_\mathrm{raw})_{f_\mathrm{clin},f_\mathrm{msr}}$ (single det.) & & & & \\
Positioning system/FS center $w_x$/$w_y$ & B & & 0.1/0.1~mm & \\
Positioning system/FS center & B & 0.02\% & 0.01\% & - \\
Readout/Machine reproducibility & A & 0.1\% & 0.1\% & 0.1\% \\
Jaws reproducibility $FS_x$/$FS_y$ & A & & 0.02/0.05~mm & \\
Jaws reproducibility $S_\mathrm{clin}$ & A & 0.026~mm~(0.32\%) & 0.027~mm~(0.08\%) & 0.027~mm~(0.02\%) \\
Combined uncertainty (k=1) & & 0.34\% & 0.13\% & 0.10\% \\
\hline
Field size determination & & & & \\
Statistical $FS_x$/$FS_y$ (1 centering) & A & & 0.06/0.07~mm & \\
Statistical $FS_x$/$FS_y$ (2 centerings) & A & & 0.04/0.05~mm & \\
Inter-det-type $FS_x$/$FS_y$ & B & & 0.07/0.08~mm & \\
$S_\mathrm{clin}$ (1 centering)$\square$ & A & 0.071~mm~(0.85\%) & 0.071~mm~(0.21\%) & 0.071~mm~(0.04\%) \\
$S_\mathrm{clin}$ (2 centerings)$\star$ & A & 0.062~mm~(0.75\%) & 0.062~mm~(0.18\%) & 0.062~mm~(0.04\%) \\
\hline
Combined uncertainty (k=1) $\square$ & &  1.75\% & 0.74\% & 0.53\% \\ 
Combined uncertainty (k=1) $\star$ & &  1.08\% & 0.61\% & 0.52\% \\ 
\hline
\hline
\end{tabular}
\label{tab:unc_budget_kfclin_mD}
}
\end{table}

\begin{table}
\centering

\caption{Uncertainty budget example for $\OF$ measurements with the PTW60019
detector. The sources of uncertainty of PTW60019 $\kfclin$,
$M_{f_\mathrm{clin},f_\mathrm{msr}}$ and field determination are explicitly detailed. The
$\star$  denotes uncertainties from two detectors with four measurements each,
and the $\square$ indicates uncertainty from a single measurement with one
detector. }

{\footnotesize
\begin{tabular}{>{\arraybackslash}m{0.42\textwidth}>
{\centering}p{0.06\textwidth}>
{\centering}p{0.15\textwidth}>
{\centering}p{0.15\textwidth}>
{\centering\arraybackslash}p{0.15\textwidth}}
\hline\hline
\multicolumn{1}{c}{\multirow{2}{*}{Item}} & \multirow{2}{*}{Type} & \multicolumn{3}{c}{Field size {[}cm²{]}} \\
\multicolumn{1}{c}{} & & 0.6 $\times$ 0.6 & 1 $\times$ 1 & 2 $\times$ 2 \\
\hline
$\kfclin$ PTW60019 (see table \ref{tab:unc_budget_kfclin_mD}) $\square$ $\star$ & B & 1.08\% & 0.61\% & 0.52\% \\
\hline
$M_{f_{clin},f_{msr}}$ PTW60019 & & & & \\ 
\hspace{8 pt}Statistical $M_{raw}$ (1 meas) $\square$ & A & 1.3\% & 0.42\% & 0.10\% \\ 
\hspace{8 pt}Statistical $M_{raw}$ (2 dets $\times$ 4 meas) $\star$ & A & 0.46\% & 0.15\% & 0.04\% \\ 
\hspace{8 pt}$\kion$ $\square$ $\star$ & B & -\% & -\% & -\% \\ 
\hspace{8 pt}$k_{stem}$ $\square$ $\star$ & B & 0.1\% & 0.1\% & 0.1\% \\ 
\hspace{8 pt}$k_{other}$ $\square$ $\star$ & B & 0.1\% & 0.1\% & 0.1\% \\ 
\hspace{8 pt}Statistical $k_{pos}$ (2 dets) $\square$ $\star$ & A & - & - & - \\ 
\hspace{8 pt}Geometry $k_{pos}$ $\square$ $\star$ & B & - & - & - \\
\hline
Field size determination & & & & \\
\hspace{8 pt}$S_{clin}$ (1 centering) $\square$ $\star$ & A & 0.85\% & 0.21\% & 0.04\% \\
\hline
Combined uncertainty (k=1) $\square$ & &  1.90\% & 0.78\% & 0.55\% \\ 
Combined uncertainty (k=1) $\star$ & &  1.46\% & 0.68\% & 0.54\% \\ 
\hline
\hline
\end{tabular}
\label{tab:unc_budget_OF_mD}
}
\end{table}

%% file: content/results/res_kfclin_dets.tex
\subsection{$\kfclin$ of all detectors}

\begin{figure}
    \centering
    \includegraphics[width=0.7\textwidth]{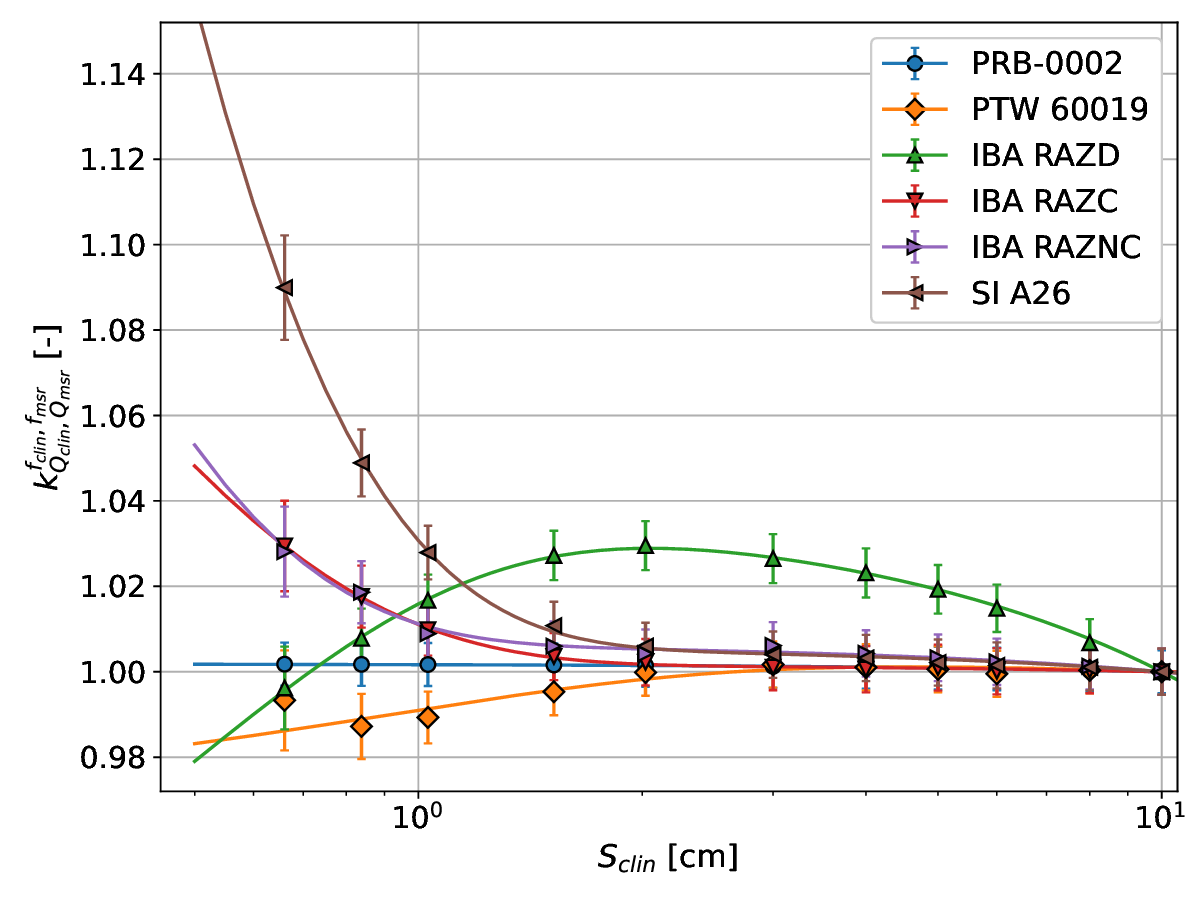}
    \caption{Extracted output correction factors $\kfclin$ for different detectors as a function of field size. The fit combines a modified sigmoidal component and a linear adjustment to model both saturating behavior and subsequent linear trends.} 
    \label{fig:kfclin_detectors}
\end{figure}

Field output correction factors evaluated in this study for all detectors
irradiated with 6 MV Varian Truebeam fields of field sizes in the range of
0.6~$\times$~0.6~\si{\centi\meter\squared} to
40~$\times$~40~\si{\centi\meter\squared} are presented in
Figure~\ref{fig:kfclin_detectors} and Table~\ref{tab:kfclin_detectors_chuq}.
Comparisons with published values are shown in Supplementary material
section~\ref{ap:kfclin_comp}. Considering uncertainties, $\kfclin$ of \probe{}
is the closest to unity of all studied detectors. Uncertainties on correction
factors obtained with the \probe{} are lower than those listed in TRS-483
\cite{international2017iaea} and those from Casar et
al.\cite{Casar.2019,Casar.2020} for other detectors. Correction factors remain
small ($<$1.5\%) over the whole set of investigated field sizes for the PTW
microDiamond. IBA Razor diode exhibits known higher correction factors due to an
increased response to lower energy photons present in larger fields. Due to the
$msr$ field normalization, IBA RAZD field output correction factors range from
$+3\%$ for 2~$\times$~2~\si{\centi\meter\squared} down to $-10\%$ at
40~$\times$~40~\si{\centi\meter\squared}. Micro chambers exhibit a
non-negligible correction factor ($>$1\%) for field sizes of
1~$\times$~1~\si{\centi\meter\squared} and below.

Comparison of $\kfclin$ with published data from
TRS-483~\cite{international2017iaea} shows only a statistically significant
difference for the PTW microDiamond at field sizes of
1~$\times$~1~\si{\centi\meter\squared} and below. Since published data from
Casar et al.~\cite{Casar.2019,Casar.2020} applies specifically to a total
correction factor and not directly to $\kfclin$, we divided their reported total
correction factor by our interpolated $\kpol$ and $\kion$ ratios between $clin$
and $msr$ fields, in order to be able to compare $\kfclin$ values with the same
definition. No statistically significant differences were then observed on the
whole field size range. Comparison with published data from Gul et
al.~\cite{gul2020} and Looe et al.~\cite{Looe2018} show however important
differences for field output correction factors of IBA Razor diodes and Nano
Chambers at small field sizes.

\input{Tables/kfclin_table_chuq}

%% file: Tables/kfclin_table_chuq.tex
\definecolor{gray92}{rgb}{0.92,0.92,0.92}
\newcolumntype{g}{>{\columncolor{gray92}}c}
\begin{table}[h]
\caption{Extracted field output correction factors for several detectors used in this study. Uncertainties are shown in brackets and represent absolute uncertainties in the last or two last digits.}
\centering
\resizebox{\linewidth}{!}{%
\begin{tabular}{cc g c g c g c}
\toprule
\toprule
\multicolumn{2}{c}{Field Size (cm)} & \multicolumn{6}{c}{$\kfclin$ (-)} \\
\cmidrule{3-8}
Nominal & Measured & \probe & PTW 60019 & IBA RAZD & IBA RAZC & IBA RAZNC & SI A26 \\
\midrule
0.60 & 0.66 & 1.002 (5) & 0.993 (11) & 0.996 (9) & 1.029 (10) & 1.028 (10) & 1.090 (12) \\
0.80 & 0.84 & 1.002 (5) & 0.987 (7) & 1.008 (6) & 1.018 (7) & 1.019 (7) & 1.049 (7) \\
1.00 & 1.03 & 1.002 (5) & 0.989 (6) & 1.017 (6) & 1.010 (6) & 1.009 (6) & 1.028 (6) \\
1.50 & 1.52 & 1.002 (5) & 0.995 (5) & 1.027 (5) & 1.004 (5) & 1.006 (5) & 1.011 (5) \\
2.00 & 2.02 & 1.001 (5) & 1.000 (5) & 1.030 (5) & 1.002 (5) & 1.004 (5) & 1.006 (5) \\
3.00 & 3.00 & 1.001 (5) & 1.002 (5) & 1.026 (5) & 1.001 (5) & 1.006 (5) & 1.004 (5) \\
4.00 & 4.00 & 1.001 (5) & 1.001 (5) & 1.023 (5) & 1.001 (5) & 1.004 (5) & 1.003 (5) \\
5.00 & 5.00 & 1.001 (5) & 1.001 (5) & 1.019 (5) & 1.001 (5) & 1.003 (5) & 1.002 (5) \\
6.00 & 6.00 & 1.001 (4) & 1.000 (5) & 1.015 (5) & 1.000 (5) & 1.002 (5) & 1.001 (5) \\
8.00 & 8.00 & 1.000 (5) & 1.000 (5) & 1.007 (5) & 1.000 (5) & 1.001 (5) & 1.001 (5) \\
10.00 & 10.00 & 1.000 (5) & 1.000 (5) & 1.000 (5) & 1.000 (5) & 1.000 (5) & 1.000 (5) \\
20.00 & 20.00 & 0.999 (5) & 0.998 (5) & 0.960 (5) & 1.000 (5) & 0.999 (5) & 1.000 (5) \\
30.00 & 30.00 & 0.999 (5) & 1.000 (6) & 0.928 (6) & 1.003 (6) & 1.000 (6) & 1.002 (6) \\
40.00 & 40.00 & 1.000 (6) & 1.002 (6) & 0.906 (6) & 1.007 (6) & 1.004 (6) & 1.006 (6) \\
\bottomrule
\bottomrule
\end{tabular}%
}
\label{tab:kfclin_detectors_chuq}
\end{table}

%% file: content/discussion.tex
\section{Discussion}

Radiation therapy is increasingly reliant on small radiation fields to deliver
highly conformal treatments. To take full advantage of these complex dose
distributions, small fields must be accurately modelled. However, measurements
in these conditions are challenging because of the changes with field size in
the dose to detector vs dose to water distribution, as well as changes in
detector response with either operating conditions (polarity for example) or
changing particle energy spectra and flux. Reports detailing advances in small
field dosimetry have recently been published \cite{international2017iaea,
Das.2021bms} to offer guidance and data in order to help define a uniform
practice.

In parallel to these efforts in improving the practice of small field dosimetry,
equipment manufacturers are also developing new instruments better suited to
measure small radiation fields. One such example is Medscint's \probe{}
studied in this work. This detector was designed specifically to minimize the
correction factors needed when measuring output factors of small fields. As seen
of Fig.~\ref{fig:perturbations}, perturbations caused by the density and atomic
composition of the scintillator, $\pscint$ is nearly entirely balanced by
changes in ionization quenching, $\kioq$, therefore providing a $\kfclin$ close
to unity for the entire range of field sizes considered. Minimizing the
amplitude of correction factors that must be applied to obtain the dose in water
from a detector's reading makes this process less sensitive to errors, therefore
simplifying both the experimental procedure and dose calculation itself. 

A careful uncertainty analysis is critical for characterization of new detectors
for a meaningful, quantitative comparison with existing detectors. Throughout
this work efforts were made to keep uncertainty as small as possible. A detailed
uncertainty budget was provided that included: averaging measurements from
multiple detectors of the same model, using an 11-points positioning technique,
evaluation of jaw reproducibility and sensitivity assessment for analytical and
Monte Carlo calculations. It is important to note that, at the smallest field
size, the largest source of uncertainty comes from the field size determination.

The detailed uncertainty characterization makes it possible to identify small
factors of influence that could otherwise be attributed to stochastic
fluctuations. One such factor for PSD is quenching. While ionization quenching
in scintillators has been known for decades \cite{birks1956scintillation}, it
was generally assumed to be negligible for megavoltage photon beams. However,
Santurio and Andersen recently showed that the contribution of low-energy
secondary electrons is sufficient to cause quenching; furthermore, the changing
spectrum of secondary electrons with field size causes a dependence on the
quenching effect importance with field size~\cite{Santurio.2019}. This means
that $\kfclin$ for PSD can be farther from unity than initially
claimed~\cite{international2017iaea}. This work is the first to quantify $\kioq$
for a commercial PSD and to show how it can be used to produce a $\kfclin$ as
close to unity as possible. 

While great care was taken to express our findings within the framework
established by guiding documents~\cite{international2017iaea,Das.2021bms}, it
was not entirely possible to do so. In this work we've explicitly differentiated
two broad categories of correction for factors of influence: type~I and type~II.
Both types represent corrections that must be applied to a \emph{raw} detector
reading so that the corrected values can be assumed to be proportional to the
average dose deposited in the sensitive volume of a detector (see
eqs.~(\ref{eq:m_det1}) and~(\ref{eq:m_det2})). Type~I factors are assumed to be
independent of small beam quality changes that can occur between a reference
field and a measured field (i.e. going from $Q_\mathrm{msr}$ to
$Q_\mathrm{clin}$ does not change $k_I$ factors). They are associated with known
external factors of influence (e.g. $k_{TP}$), that can also change with field
size (e.g. $\kion$, $\kpol$), but not directly because of a beam quality change.
Following TRS-483, they are applied to raw measurement before the field output
correction factor. With type~II factors, small changes in beam quality occurring
when going from $f_\mathrm{msr}$ to $f_\mathrm{clin}$ is considered within the
cause of intrinsic detector response variation. Because these factors do not
involve an influencing quantity that is externally measurable and easily
factorizable, they are herein proposed to be included in the field output
correction factor. Traditionally, all corrections were assumed to be independent
of beam quality changes and thus fell in the type~I category. We believe the
distinction between categories I and II is important to be made explicit even
if, for now, it mainly concerns PSD detectors. Other corrections factors could
be classified as type~II. For example, a part of $\kion$ ratio is attributed to
initial recombination, which is beam quality dependent and could be seen as
type~II. With the proposed equation for $\kion$ ratio correction (see
eq.~\ref{eq:kion_ratio2}), it is assumed that the $\kion$ change with field size
is only dependent on the dose per pulse variation. Therefore, if there is any
change in initial recombination rate with field size, this will be included in
the field output correction factor. However, knowing the rather small importance
of general ion recombination variations for IC and SS detectors in the dose per
pulse range of standard electron/photon external beam radiotherapy (see
table~\ref{tab:B}), it can be expected that the remaining initial recombination
variation will be negligible.

Using eq.~\ref{eq:kfclin_psd}, field output correction factors for the new
Medscint's \probe{} were determined with an uncertainty of about 0.5\% (see
Table~\ref{tab:unc_budget_kfclin_HS} for details). This precision is comparable
or better than $\kfclin$ uncertainties of other detectors listed in
TRS-483~\cite{international2017iaea}. Furthermore, using this detector to
measure output factors (see Table~\ref{tab:unc_budget_OF_HS}) may lead to more
precise measurements than those listed in
references~\cite{international2017iaea, Casar.2019}. Mostly excellent agreement
is seen in the comparison of field output correction factors of different
detectors to published data for the same type of linear accelerator and detector
orientation~\cite{international2017iaea, Casar.2019, Casar.2020, Lechner.2022,
Mateus2023}. However, care should be taken to follow the same field output
correction factor definition. For example, Casar et al. implicitly included the
ion recombination and polarity effects in the correction factor. This had to be
factored out in order to compare $\kfclin$ values following TRS-483 definition.
It is to be noted that, as other have reported (see comments on TRS-483 by Das
and refs therein~\cite{Das.2018} and following reply~\cite{Palmans2018}), we
measured a small turnaround of $\kfclin$ for PTW 60019 at the smallest field
size that is significantly different than what is reported in TRS-483. This
effect arises at the low end of the region where volume averaging is partially
countered by the density effect of diamond over-response that changes when field
size approaches detector dimensions~\cite{Ringholz2023}. However, it could also
involve radiation-induced charge imbalance created in the connections below the
sensitive volume~\cite{Looe.2019}. The reason why some studies show this feature
and others don't is still not definitively answered. Another statistically
significant difference observed was between our measured small fields $\kfclin$
factors for the newer IBA Razor Nano chamber and the ones reported by Gul et
al.~\cite{gul2020} and Looe et al.~\cite{Looe2018} (see Supplementary material
table~\ref{tab:kfclin_detectors_gul_looe_mateus}). Despite both correcting for
polarity effect, they observe a lower field output correction factor. The fact
that the $\kfclin$ factor reported by Gul et al. for the IBA Razor diode is also
lower might indicate a difference in their reference detector output ratio. A
much better agreement for the IBA Razor Nano chamber is seen with recent Mateus
et al.~\cite{Mateus2023} data which uses a TRS-483 corrected PTW 60019 detector
as reference and with data reported by Girardi et al.~\cite{Girardi.2018P179}
with Gafchromic EBT3 films as reference. These differences between studies seen
with the smallest volume micro ionization chamber might also need further
investigations.

Measuring precise field output factors of small radiation fields can be a
rewarding process when modelling beam source parameters for a TPS, as it will
impact the quality of dose calculations for simple small target cases or more
complex modulated treatment techniques. However, great care should be deployed
in measuring and applying correction factors. Also understanding the physical
effects involved in $\kfclin$ at small fields is important to better asses
uncertainties and potential zones of detector development. As examples,
table~\ref{tab:kvol_detectors} in Supplementary material
section~\ref{ap:kvol_profiles} illustrates the importance of volume averaging
effects following the choice of detector and orientation, and Monte Carlo
studies separating cameral, sensitive material and density, as well as
recombination and signal collection effects, are useful in tackling detector
design issues and new opportunities. However, given the work involved in
determining these correction factors, day-to-day clinical practice can be
simplified by selecting a detector that required only minimal corrections. 

With \probe{}, taking advantage of having a $\kfclin$ close to unity with small
and well characterized uncertainties, we used eq.~\ref{eq:kfclin_det2} to
determine the $\kfclin$ of other detectors and then used these detectors to
measure $\OF$. For example, applying this process to the PTW microDiamond,
uncertainties between 1.9\% ($0.6\times 0.6$~cm$^2$ field) and 0.55\% ($2\times
2$~cm$^2$ field) can be reached for a single measurement.

%% file: content/conclusion.tex
\section{Conclusion}

In this work we first present a detailed determination of field output
correction factor, $\kfclin$ for the new Medscint's \probe{}; then we used that
value to determine the detector specific field output correction factors for a
large number of detectors, both ionization chambers and solid state detectors.
Our work complement and validate the results from other publications such as on
$\kpol$ \cite{Looe2018} and $\kfclin$ \cite{international2017iaea, Casar.2019,
gul2020}. We carefully considered the causes and magnitude of uncertainties and
provided realistic measurement scenarios with uncertainty budgets
(Table~\ref{tab:unc_budget_kfclin_HS} to \ref{tab:unc_budget_OF_mD}) to help
others in the challenging task of accurate dose measurement in small fields.

Ionization quenching in PSDs have been shown to have a measurable impact in
megavoltage photon beam when going from small to large fields
\cite{Santurio.2019}. This is explained by the increased amount of low energy
scatter radiation as field size becomes larger. Low energy radiation cause more
quenching and that is why $\kioq$ increase with field size. The change of energy
spectra with field size also impact $\pscint$, the perturbations due to the variation in
density and atomic composition of the scintillating material compared to water.
Our work is the first to show that for the new Medscint's \probe{}, the impact
of $\kioq$ and $\pscint$ compensate each other to produce a $\kfclin$ close to 1
for all field sizes.

%% file: content/acknowledgements.tex
\section{Acknowledgements}

The authors would like to thank Malcolm McEwen for his insightful comments on
the manuscript,  Grishar Valdes Santurio for helpful discussions on $\kioq$, and
Hugo Bouchard for useful discussions regarding the uncertainty budget. The
authors also thank Jean-François Aubry and Robert Doucet from CHUM for lending a
microDiamond detector. This research was enabled in part by the support provided
by Calcul Qu\'ebec (calculquebec.ca) and  the Digital Research Alliance of
Canada (alliancecan.ca). This work was supported by Mitacs through the Mitacs
Accelerate program. 

\section{Conflict of interests}

Some authors (BC, SL, DL, YL, FT) are employees of Medscint, which produces one
of the products discussed in this paper. They participated in the data
acquisition and Monte Carlo simulation components of the study; however, the
research and conclusions presented were conducted independently and are not
influenced by any commercial interests.

%% file: content/appendix/k_defs.tex

\section{Definitions of $k_I$ corrections}\label{ap:kdef}

\begin{itemize}
   \item $k_\mathrm{TP}$: the linear temperature and pressure corrections
   normalizing ionization chamber environmental response to standard reference
   condition (22~$^{\circ}$C, 101.33~kPa)\cite{Almond.1999}.
   \item $k_\mathrm{H}$: air humidity correction factor. For operations in the
   range of 20-80\% relative humidity, $k_\mathrm{H}$ may be considered as
   unity with a 0.15\% uncertainty \cite{Almond.1999}.
   \item $k_\mathrm{T}$: temperature correction factor normalizing SS and PSD
   detector response to standard reference condition of 22~$^{\circ}$C.
   \item $k_\mathrm{elec}$: electrometer correction factor to correctly
   account for electric current produced in the detector.
   \item $k_\mathrm{read}$: readout correction for signals from devices
   other than an electrometer. For PSDs, it is the conversion factor from raw
   digital readout of the optical sensor back to the light output of the
   detector. It can change with time and can be a result of sensor thermal drift or
   optical coupling variations.
   \item $k_\mathrm{pol}^{+/-}$: ion chamber polarity correction factor
   that averages charge accumulated at both positive and negative
   polarity and is applied to signal acquired with either $+$ or $-$ polarity.
   \item $k_\mathrm{ion}$: ionization recombination correction factor for
   ionization chambers that depends on the detector operation polarizing
   voltage \cite{Almond.1999}. For SS detectors this is the incomplete charge
   collection factor that accounts for change in sensitivity and dose rate
   dependence due to recombination-generation centres and traps included in the
   crystalline structure of the sensitive volume.
   \item $k_\mathrm{drift}$: correction for machine output drifts with time.
   \item $k_\mathrm{bg}$: correction for radiation dose contribution that do
   not come from an intended irradiation from the dose delivery apparatus. This
   correction factor should be unity if all background contributions are
   correctly subtracted from the measured raw signal.
   \item $k_\mathrm{pos}$: detector position uncertainty correction factor.
   $M^\mathrm{det}$ is intended to be the measurement corrected for all
   influence quantities, other than beam quality, but including detector
   positional uncertainty. This means that this is the value obtained with
   perfect detector positioning. This factor therefore correct for impact of
   position uncertainty. See section~\ref{sec:kvol} for more details.
   \item $k_\mathrm{stem}$: this correction factor is intended to remove any
   contribution to the raw measurement that come from irradiation of any part
   of the detector system other than the sensitive volume. Along with
   $k_\mathrm{bg}$, this is the second correction needed for signal leakage
   removal.
   \item $k_\mathrm{other}$: any other correction factor
   
 

\end{itemize}

%% file: content/appendix/det_geo-orient_integration.tex
\section{Determination of detectors geometrical and positional response to
dose distribution}\label{ap:detshape}

Detector sensitive volume geometries may be either spherical, with a radius $r$;
cylindrical, with a radius $r$ and length $l$; or cylindrical with a
half-spherical tip, with a cylindrical length $l$ and common sphere and cylinder
radius $r$. Two orientations were considered for cylindrical detectors, with
either their symmetry axis parallel or perpendicular to the beam axis (along $z$
or $y$ direction respectively). 

Given these geometries, the integral in eq.~(\ref{eq:mxy}) is performed either
in a Polar or Cartesian coordinate system, with the coordinate transformations
$x-x_0 = \rho \cos \theta$ and $y-y_0 = \rho \sin \theta$. Table~\ref{tab:hfunc}
presents the coordinate system, analytical function $h$, and cross sectional
area $A$ used for each combination of geometry and orientation.
Table~\ref{tab:mxy} gives the functions $M(x_0,y_0)$ obtained from the
integration of eq.~(\ref{eq:mxy}). The expected normalized measurement signal functions,
$\left\langle M(x_0,y_0)\right\rangle$, obtained from eq.~(\ref{eq:expmxy})
using the equations from Table~\ref{tab:mxy} are given in Table~\ref{tab:expmxy}. 

\begin{table}[ht]
\centering 
\caption{Detector height function and cross-sectional area used in equation~(\ref{eq:mxy}).}
\resizebox{\linewidth}{!}{%
\begin{tabular} {|c|c|c|c|c|}
\hline\hline
Geometry & Orientation  & Coord. System & $h$ & $A$ \\
\hline
Spherical & - & Polar & $\frac{3}{2 \pi r^3} \sqrt{r^2 - \rho^2}$ & $\int_{0}^{r} \int_{0}^{2\pi} \rho d\theta d\rho$ \\
Cylindrical & $\|$ z & Polar & $\frac{1}{\pi r^2}$ & $\int_{0}^{r} \int_{0}^{2\pi} \rho d\theta d\rho$ \\
Cylindrical & $\perp$ y & Cartesian & $\frac{2}{\pi r^2 l} \sqrt{r^2 - (x-x_0)^2}$ & $\int_{-l/2}^{l/2} \int_{-r}^{r} dx dy$ \\
Cyl / Half-Sph & $\|$ z & Polar & $\frac{1}{\pi r^2 (l + 2r/3)} (l + \sqrt{r^2 - \rho^2})$ & $\int_{0}^{r} \int_{0}^{2\pi} \rho d\theta d\rho$ \\
\hline\hline
\end{tabular}%
}
\label{tab:hfunc}
\end{table}

\begin{table}[ht]
\centering 
\caption{Analytical formulation of normalized measurement signal described by
eq.~(\ref{eq:mxy}) at position ($x_0,y_0$) for different detector geometry and
orientation.} \resizebox{0.8\linewidth}{!}{%
\begin{tabular} {|c|c|c|}
\hline\hline
Geometry & Orientation  & $M(x_0,y_0)$ \\
\hline
Spherical & - & $(f(x_0)+\frac{a_2r^2}{5})(g(y_0)+\frac{b_2r^2}{5}) - \frac{a_2b_2r^4}{70}$ \\
Cylindrical & $\|$ z & $(f(x_0)+\frac{a_2r^2}{4})(g(y_0)+\frac{b_2r^2}{4}) - \frac{a_2b_2r^4}{48}$ \\
Cylindrical & $\perp$ y & $(f(x_0)+\frac{a_2r^2}{4})(g(y_0)+\frac{b_2r^2}{12})$ \\
Cyl / Half-Sph & $\|$ z & $\frac{2r/3}{l+2r/3}M_{sph}(x_0,y_0) + \frac{l}{l+2r/3}M_{cyl\parallel}(x_0,y_0)$ \\
\hline\hline
\end{tabular}%
}
\label{tab:mxy}
\end{table}

\begin{table}[ht]
\centering 
\caption{Analytical formulation of expected normalized measurement signal, $M$,
given by eq.~(\ref{eq:expmxy}) at position ($x_0,y_0$) for different detector
geometries and orientations.} \resizebox{\linewidth}{!}{%
\begin{tabular} {|c|c|c|}
\hline\hline
Geometry & Orientation & $\left\langle M(x_0,y_0) \right\rangle$ \\
\hline
Spherical & - & $M(x_0,y_0) + (f(x_0)+\frac{a_2r^2}{5})\frac{b_2w_y^2}{3} + (g(y_0)+\frac{b_2r^2}{5})\frac{a_2w_x^2}{3} + \frac{a_2b_2w_x^2w_y^2}{9}$ \\
Cylindrical & $\|$ z & $M(x_0,y_0) + (f(x_0)+\frac{a_2r^2}{4})\frac{b_2w_y^2}{3} + (g(y_0)+\frac{b_2r^2}{4})\frac{a_2w_x^2}{3} + \frac{a_2b_2w_x^2w_y^2}{9}$ \\
Cylindrical & $\perp$ y & $M(x_0,y_0) + (f(x_0)+\frac{a_2r^2}{4})\frac{b_2w_y^2}{3} + (g(y_0)+\frac{b_2r^2}{12})\frac{a_2w_x^2}{3} + \frac{a_2b_2w_x^2w_y^2}{9}$ \\
Cyl / Half-Sph & $\|$ z & $\frac{2r/3}{l+2r/3}\left\langle M_{sph}\right\rangle(x_0,y_0) + \frac{l}{l+2r/3}\left\langle M_{cyl\parallel} \right\rangle (x_0,y_0)$ \\
\hline\hline
\end{tabular}%
}
\label{tab:expmxy}
\end{table}

%% file: content/appendix/kvol_psd_all_profiles.tex
\section{Impact of the measured profiles on $\kvol$ for \probe{}}\label{ap:kvol_profiles}

Table~\ref{tab:kvol_detectors} shows $kvol$ correction factors for each detector
determined with their respective fitted profiles and appropriate geometrical
functions (see Eq.~\ref{eq:dxy} to \ref{eq:kvol_def} and Supplementary material
section~\ref{ap:kvol_profiles}). Because profiles can be measured with any
detector, Table~\ref{tab:kvol_hs1mm} compares impact of using different
detectors for profile measurements to determine $\kvol$. The average value was
used to determine $\kfclin$ of Medscint's \probe{}.

\input{Tables/kvol_det_table}
\input{Tables/kvol_hs1mm_table}


%% file: Tables/kvol_det_table.tex
\definecolor{gray92}{rgb}{0.92,0.92,0.92}
\newcolumntype{g}{>{\columncolor{gray92}}c}
\begin{table}[h]
\caption{Extracted volume component of field output correction factors. All detectors aligned parallel to beam axis, except Medscint PSD which is perpendicular along the inline axis.}
\centering
\resizebox{\linewidth}{!}{%
\begin{tabular}{cc g c g c g c}
\toprule
\toprule
\multicolumn{2}{c}{Field Size (cm)} & \multicolumn{6}{c}{$k_{vol}$ (-)} \\
\cmidrule{3-8}
Nominal & Measured & \probe{} ($\perp$) & PTW 60019 & IBA RAZD & IBA RAZC & IBA RAZNC & SI A26 \\
\midrule
0.50 & 0.54 & - & 1.033 & 1.003 & 1.026 & 1.025 & - \\
0.60 & 0.64 & 1.006 & 1.023 & 1.002 & 1.021 & 1.017 & 1.040 \\
0.80 & 0.82 & 1.003 & 1.012 & 1.001 & 1.011 & 1.012 & 1.028 \\
1.00 & 1.00 & 1.001 & 1.007 & 1.000 & 1.005 & 1.005 & 1.014 \\
2.00 & 1.99 & - & 1.000 & 1.000 & 1.001 & 1.000 & 1.001 \\
\bottomrule
\bottomrule
\end{tabular}%
}
\label{tab:kvol_detectors}
\end{table}

%% file: Tables/kvol_hs1mm_table.tex
\definecolor{gray92}{rgb}{0.92,0.92,0.92}
\newcolumntype{g}{>{\columncolor{gray92}}c}
\begin{table}[h]
\caption{$\kvol$ for \probe{} determined using dose profiles measured with different detectors, with detector axis either parallel ($\parallel$) or perpendicular ($\perp$) to the photon beam axis.}
\centering
\resizebox{\linewidth}{!}{%
\begin{tabular}{cc gg cc gg cc gg cc gg}
\toprule
\toprule
\multicolumn{2}{c}{} & \multicolumn{14}{c}{$k_{vol}^{PSD}$ (-)} \\
\cmidrule{3-16}
\multicolumn{2}{c}{Field Size (cm)} & \multicolumn{2}{g}{\probe{}} & \multicolumn{2}{c}{PTW 60019} & \multicolumn{2}{g}{IBA RAZD} & \multicolumn{2}{c}{IBA RAZC} & \multicolumn{2}{g}{IBA RAZNC} & \multicolumn{2}{c}{SI A26} & \multicolumn{2}{g}{Average} \\
Nominal & Measured & $\parallel$ & $\perp$ & $\parallel$ & $\perp$ & $\parallel$ & $\perp$ & $\parallel$ & $\perp$ & $\parallel$ & $\perp$ & $\parallel$ & $\perp$ & $\parallel$ & $\perp$ \\
\midrule
0.50 & 0.54 & - & - & 1.0066 (6) & 1.0077 (4) & 1.0073 (7) & 1.0085 (5) & 1.0076 (7) & 1.0089 (5) & 1.0078 (8) & 1.0091 (5) & - & - & 1.0073 (3) & 1.0086 (2) \\
0.60 & 0.64 & 1.0050 (5) & 1.0058 (2) & 1.0047 (4) & 1.0054 (2) & 1.0047 (4) & 1.0055 (3) & 1.0063 (6) & 1.0073 (3) & 1.0052 (5) & 1.0060 (3) & 1.0045 (4) & 1.0052 (2) & 1.0050 (2) & 1.0059 (1) \\
0.80 & 0.82 & 1.0022 (2) & 1.0026 (1) & 1.0025 (2) & 1.0029 (1) & 1.0025 (2) & 1.0030 (1) & 1.0032 (3) & 1.0037 (2) & 1.0038 (3) & 1.0044 (2) & 1.0031 (3) & 1.0036 (1) & 1.0029 (1) & 1.0034 (0) \\
1.00 & 1.00 & 1.0011 (1) & 1.0013 (0) & 1.0013 (1) & 1.0016 (0) & 1.0012 (1) & 1.0014 (0) & 1.0016 (1) & 1.0019 (1) & 1.0016 (1) & 1.0019 (0) & 1.0016 (1) & 1.0019 (0) & 1.0014 (0) & 1.0017 (0) \\
2.00 & 1.99 & - & - & 1.0001 (0) & 1.0001 (0) & 1.0001 (0) & 1.0001 (0) & 1.0002 (0) & 1.0002 (0) & 1.0001 (0) & 1.0002 (0) & 1.0001 (0) & 1.0001 (0) & 1.0001 (0) & 1.0001 (0) \\
\bottomrule
\bottomrule
\end{tabular}%
}
\label{tab:kvol_hs1mm}
\end{table}

%% file: content/appendix/kfclin_comp.tex
\section{Comparisons of $\kfclin$ with published values}\label{ap:kfclin_comp}

Values for $\kfclin$ obtained in this study are compared with other published
values. Table~\ref{tab:kfclin_detectors_trs483} shows the comparison with values
from the TRS-483 report \cite{international2017iaea},
Table~\ref{tab:kfclin_detectors_casar_corr} compares with the works of Casar et
al.\cite{Casar.2019, Casar.2020} and
Table~\ref{tab:kfclin_detectors_gul_looe_mateus} compares with Gul et
al.\cite{gul2020}, Looe et al.~\cite{Looe2018} and Mateus et
al.~\cite{Mateus2023}. In these tables, values of $\kfclin$ from this study are
extracted from our fit equation at the same reported $S_\mathrm{clin}$ field size as
the compared literature data. Please note that for TRS-483 the $\kfclin$ values
for ion chambers are tabulated for perpendicular orientations while our ion
chamber measurements were done with parallel orientations. This should not
impact A26 which is spherical, but could have a minor effect on the RAZC at
smallest fields.

\input{Tables/kfclin_table_comp_trs483}
\input{Tables/kfclin_table_comp_casar_corr}
\input{Tables/kfclin_table_comp_gul_looe_mateus}

%% file: Tables/kfclin_table_comp_trs483.tex
\definecolor{gray92}{rgb}{0.92,0.92,0.92}
\newcolumntype{g}{>{\columncolor{gray92}}c}
\begin{table}[h]
\caption{Comparison of field output correction factors from this study and values from IAEA TRS-483 \cite{international2017iaea} for several detectors. Uncertainties are shown in brackets and represent absolute uncertainties in the last or two last digits.}
\centering
\resizebox{\linewidth}{!}{%
\begin{tabular}{c ggg ccc ggg ccc}
\toprule
\toprule
& \multicolumn{12}{c}{$\kfclin$ (-)} \\
\cmidrule{2-13}
& \multicolumn{3}{g}{PTW 60019} & \multicolumn{3}{c}{IBA RAZD/SFD} & \multicolumn{3}{g}{IBA RAZC/CC01} & \multicolumn{3}{c}{SI A26}\\
Field Size (cm) & this study & TRS-483 & p-value & this study & TRS-483 & p-value & this study & TRS-483 & p-value & this study & TRS-483 & p-value \\
\midrule
0.60 & 0.985 (8) & 0.968 (7) & 0.001 & 0.990 (11) & 0.990 (6) & 0.998 & 1.035 (14) & 1.047 (26) & 0.391 & 1.110 (16) & 1.165 (29) & 0.188 \\
0.80 & 0.988 (6) & 0.977 (5) & 0.002 & 1.006 (6) & 1.007 (6) & 0.747 & 1.019 (7) & 1.027 (16) & 0.352 & 1.056 (7) & 1.062 (16) & 0.714 \\
1.00 & 0.991 (5) & 0.984 (4) & 0.014 & 1.016 (5) & 1.018 (5) & 0.446 & 1.011 (5) & 1.018 (11) & 0.230 & 1.031 (5) & 1.023 (11) & 0.502 \\
1.50 & 0.996 (4) & 0.993 (4) & 0.276 & 1.027 (4) & 1.030 (5) & 0.227 & 1.003 (4) & 1.011 (6) & 0.061 & 1.010 (4) & 1.003 (6) & 0.324 \\
2.00 & 0.998 (4) & 0.997 (3) & 0.574 & 1.029 (5) & 1.032 (4) & 0.243 & 1.002 (4) & 1.009 (4) & 0.055 & 1.006 (5) & 1.000 (4) & 0.248 \\
3.00 & 1.000 (4) & 1.000 (4) & 0.821 & 1.027 (5) & 1.029 (4) & 0.395 & 1.001 (5) & 1.008 (4) & 0.069 & 1.004 (5) & 1.000 (4) & 0.361 \\
4.00 & 1.001 (4) & 1.000 (3) & 0.569 & 1.023 (4) & 1.025 (3) & 0.407 & 1.001 (4) & 1.007 (4) & 0.082 & 1.004 (4) & 1.000 (4) & 0.415 \\
5.00 & 1.001 (4) & 1.000 (3) & 0.522 & 1.019 (4) & 1.021 (3) & 0.390 & 1.001 (4) & 1.006 (4) & 0.121 & 1.003 (3) & 1.000 (4) & 0.478 \\
6.00 & 1.001 (3) & 1.000 (3) & 0.533 & 1.015 (3) & 1.017 (3) & 0.357 & 1.001 (3) & 1.004 (3) & 0.159 & 1.002 (2) & 1.000 (3) & 0.455 \\
8.00 & 1.001 (1) & 1.000 (3) & 0.654 & 1.008 (1) & 1.008 (3) & 0.792 & 1.000 (1) & 1.002 (3) & 0.299 & 1.001 (1) & 1.000 (3) & 0.677 \\
10.00 & 1.000 (0) & 1.000 (0) & 1.000 & 1.000 (0) & 1.000 (0) & 1.000 & 1.000 (0) & 1.000 (0) & 1.000 & 1.000 (0) & 1.000 (0) & 1.000 \\
\bottomrule
\bottomrule
\end{tabular}%
}
\label{tab:kfclin_detectors_trs483}
\end{table}

%% file: Tables/kfclin_table_comp_casar_corr.tex
\definecolor{gray92}{rgb}{0.92,0.92,0.92}
\newcolumntype{g}{>{\columncolor{gray92}}c}
\begin{table}[h]
\caption{Comparison of field output correction factors from this study and values from Casar et al.~\cite{Casar.2019, Casar.2020} for three detectors. ($*$): Values from reference, but modified to exclude expected $\kpol$ and $\kion$ factors. Uncertainties are shown in brackets and represent absolute uncertainties in the last or two last digits.}
\centering
\resizebox{\linewidth}{!}{%
\begin{tabular}{c ggg ccc ggg}
\toprule
\toprule
& \multicolumn{9}{c}{$\kfclin$ (-)} \\
\cmidrule{2-10}
& \multicolumn{3}{g}{PTW 60019} & \multicolumn{3}{c}{IBA RAZD} & \multicolumn{3}{g}{IBA RAZC} \\
Field Size (cm) & this study & Casar et al.~\cite{Casar.2019}$^{*}$ & p-value & this study & Casar et al.~\cite{Casar.2019}$^{*}$ & p-value & this study & Casar et al.~\cite{Casar.2020}$^{*}$ & p-value \\
\midrule
0.56 & 0.984 (9) & 0.976 (19) & 0.515 & 0.986 (13) & 0.991 (19) & 0.731 & 1.040 (17) & 1.063 (22) & 0.217 \\
0.81 & 0.989 (6) & 0.970 (13) & 0.122 & 1.007 (6) & 1.000 (13) & 0.494 & 1.019 (6) & 1.018 (14) & 0.937 \\
1.01 & 0.991 (5) & 0.985 (12) & 0.488 & 1.016 (5) & 1.018 (12) & 0.794 & 1.011 (5) & 1.016 (12) & 0.529 \\
1.50 & 0.996 (4) & 0.992 (11) & 0.669 & 1.027 (4) & 1.023 (11) & 0.595 & 1.003 (4) & 1.004 (11) & 0.966 \\
2.00 & 0.998 (4) & 0.995 (11) & 0.712 & 1.029 (5) & 1.023 (12) & 0.479 & 1.002 (4) & 0.999 (12) & 0.760 \\
3.03 & 1.001 (4) & 1.000 (11) & 0.896 & 1.027 (5) & 1.024 (11) & 0.733 & 1.001 (5) & 1.002 (11) & 0.879 \\
4.03 & 1.001 (4) & 0.999 (11) & 0.741 & 1.023 (4) & 1.019 (11) & 0.627 & 1.001 (4) & 1.001 (11) & 0.989 \\
5.02 & 1.001 (4) & 0.996 (10) & 0.450 & 1.019 (4) & 1.012 (11) & 0.401 & 1.001 (4) & 0.997 (10) & 0.632 \\
10.03 & 1.000 (0) & 1.001 (0) & 0.000 & 1.000 (0) & 1.001 (1) & 0.192 & 1.000 (0) & 1.001 (0) & 0.000 \\
\bottomrule
\bottomrule
\end{tabular}%
}
\label{tab:kfclin_detectors_casar_corr}
\end{table}

%% file: Tables/kfclin_table_comp_gul_looe_mateus.tex
\definecolor{gray92}{rgb}{0.92,0.92,0.92}
\newcolumntype{g}{>{\columncolor{gray92}}c}
\begin{table}[h]
\caption{Comparison of field output correction factors from this study and values from Gul et al.~\cite{gul2020}, Looe et al.~\cite{Looe2018} and Mateus et al.~\cite{Mateus2023} for two detectors. Uncertainties are shown in brackets and represent absolute uncertainties in the last or two last digits.}
\centering
\resizebox{\linewidth}{!}{%
\begin{tabular}{c ggg cccg cc}
\toprule
\toprule
& \multicolumn{9}{c}{$\kfclin$ (-)} \\
\cmidrule{2-10}
& \multicolumn{3}{g}{IBA RAZD} & \multicolumn{6}{c}{IBA RAZNC} \\
Field Size (cm) & this study & Gul et al.~\cite{gul2020} & p-value & this study & Gul et al.~\cite{gul2020} & p-value & Looe et al.~\cite{Looe2018} & Mateus et al.~\cite{Mateus2023} & p-value \\
\midrule
0.52 & 0.981 (15) & 0.948 (2) & 0.003 & 1.049 (21) & 0.998 (4) & 0.017 & 0.960 (23) & - (-) & - \\
0.60 & 0.990 (11) & - (-) & - & 1.036 (13) & - (-) & - & 0.962 (23) & 1.042 (23) & 0.724 \\
0.80 & 1.006 (6) & - (-) & - & 1.019 (7) & - (-) & - & 0.980 (23) & 1.009 (23) & 0.550 \\
1.00 & 1.016 (5) & 1.004 (4) & 0.002 & 1.011 (5) & 0.981 (4) & 0.000 & 0.991 (23) & 1.005 (23) & 0.689 \\
1.50 & 1.027 (4) & 1.019 (3) & 0.012 & 1.006 (4) & 0.991 (2) & 0.005 & 0.997 (23) & 1.000 (13) & 0.502 \\
2.00 & 1.029 (5) & 1.020 (4) & 0.011 & 1.005 (4) & 0.996 (3) & 0.023 & 1.006 (23) & - (-) & - \\
2.51 & 1.028 (5) & 1.023 (3) & 0.078 & 1.005 (4) & 0.998 (3) & 0.058 & - (-) & 1.000 (13) & 0.586 \\
3.00 & 1.027 (5) & 1.020 (2) & 0.027 & 1.005 (4) & 0.998 (2) & 0.063 & 1.001 (23) & 1.000 (13) & 0.609 \\
4.00 & 1.023 (4) & 1.018 (3) & 0.068 & 1.004 (4) & 0.997 (3) & 0.040 & 1.000 (23) & 1.000 (13) & 0.657 \\
10.00 & 1.000 (0) & 1.000 (4) & 1.000 & 1.000 (0) & 1.000 (4) & 1.000 & - (-) & 1.000 (3) & 1.000 \\
\bottomrule
\bottomrule
\end{tabular}%
}
\label{tab:kfclin_detectors_gul_looe_mateus}
\end{table}

%% file: main.bbl
\begin{thebibliography}{10}

\bibitem{lechner-2018}
W.~Lechner, P.~Wesolowska, G.~Azangwe, M.~Arib, V.~G.~L. Alves, L.~Suming,
  D.~Ekendahl, W.~Bulski, J.~L.~A. Samper, S.~P. Vinatha, S.~Siri, M.~Tomsej,
  M.~Tenhunen, J.~Povall, S.~F. Kry, D.~S. Followill, D.~I. Thwaites, D.~Georg,
  and J.~Izewska,
\newblock A multinational audit of small field output factors calculated by
  treatment planning systems used in radiotherapy,
\newblock Physics and Imaging in Radiation Oncology {\bf 5}, 58--63 (2018).

\bibitem{das-2008}
I.~J. Das, G.~X. Ding, and A.~Ahnesjö,
\newblock Small fields: nonequilibrium radiation dosimetry,
\newblock Medical Physics {\bf 35}, 206--215 (2008).

\bibitem{alfonso.2008}
R.~Alfonso, P.~Andreo, R.~Capote, M.~S. Huq, W.~Kilby, P.~Kjäll, T.~R. Mackie,
  H.~Palmans, K.~Rosser, J.~Seuntjens, W.~Ullrich, and S.~Vatnitsky,
\newblock A new formalism for reference dosimetry of small and nonstandard
  fields: {Reference} dosimetry of small and nonstandard fields,
\newblock Medical Physics {\bf 35}, 5179--5186 (2008).

\bibitem{international2017iaea}
IAEA,
\newblock {\em {Dosimetry of Small Static Fields Used in External Beam
  Radiotherapy}},
\newblock Number 483 in Technical Reports Series, INTERNATIONAL ATOMIC ENERGY
  AGENCY, Vienna, 2017.

\bibitem{Das.2021bms}
I.~J. Das, P.~Francescon, J.~M. Moran, A.~Ahnesjö, M.~M. Aspradakis, C.~Cheng,
  G.~X. Ding, J.~D. Fenwick, M.~S. Huq, M.~Oldham, C.~S. Reft, and O.~A. Sauer,
\newblock Report of AAPM Task Group 155: megavoltage photon beam dosimetry in
  small fields and non-equilibrium conditions,
\newblock Medical Physics {\bf 48}, e886--e921 (2021).

\bibitem{Casar.2019}
B.~Casar, E.~Gershkevitsh, I.~Mendez, S.~Jurković, and M.~S. Huq,
\newblock A novel method for the determination of field output factors and
  output correction factors for small static fields for six diodes and a
  microdiamond detector in megavoltage photon beams,
\newblock Medical Physics {\bf 46}, 944--963 (2019).

\bibitem{mcgrath-2022}
A.~N. McGrath, S.~Golmakani, and T.~J. Williams,
\newblock Determination of correction factors in small {MLC}-defined fields for
  the {Razor} and {microSilicon} diode detectors and evaluation of the
  suitability of the {IAEA} {TRS}-483 protocol for multiple detectors,
\newblock Journal of Applied Clinical Medical Physics {\bf 23}, e13657 (2022).

\bibitem{Casar.2020}
B.~Casar, E.~Gershkevitsh, I.~Mendez, S.~Jurković, and M.~S. Huq,
\newblock Output correction factors for small static ﬁelds in megavoltage
  photon beams for seven ionization chambers in two orientations —
  perpendicular and parallel,
\newblock Medical Physics {\bf 47}, 242--259 (2020).

\bibitem{gagnon-2011}
J.-C. Gagnon, D.~Thériault, M.~Guillot, L.~Archambault, S.~Beddar, L.~Gingras,
  and L.~Beaulieu,
\newblock Dosimetric performance and array assessment of plastic scintillation
  detectors for stereotactic radiosurgery quality assurance,
\newblock Medical Physics {\bf 39}, 429--436 (2011).

\bibitem{Papaconstadopoulos.2017}
P.~Papaconstadopoulos, L.~Archambault, and J.~Seuntjens,
\newblock Experimental investigation on the accuracy of plastic scintillators
  and of the spectrum discrimination method in small photon fields,
\newblock Medical Physics {\bf 44}, 654--664 (2017).

\bibitem{beddar-1992a}
A.~S. Beddar, T.~R. Mackie, and F.~H. Attix,
\newblock Water-equivalent plastic scintillation detectors for high-energy beam
  dosimetry: {I}. {Physical} characteristics and theoretical considerations,
\newblock Physics in Medicine and Biology {\bf 37}, 1883--1900 (1992).

\bibitem{Wang.2011}
L.~L.~W. Wang and S.~Beddar,
\newblock Study of the response of plastic scintillation detectors in
  small‐field 6 MV photon beams by Monte Carlo simulations,
\newblock Medical Physics {\bf 38}, 1596--1599 (2011).

\bibitem{Santurio.2019}
G.~V. Santurio and C.~E. Andersen,
\newblock Quantifying the ionization quenching effect in organic plastic
  scintillators used in MV photon dosimetry,
\newblock Radiation Measurements {\bf 129}, 106200 (2019).

\bibitem{Santurio.2020}
G.~V. Santurio, M.~Pinto, and C.~E. Andersen,
\newblock {Evaluation of the ionization quenching effect in an organic plastic
  scintillator using kV x-rays and a modified Birks model with explicit account
  of secondary electrons},
\newblock Radiation Measurements {\bf 131}, 106222 (2020).

\bibitem{Uijtewaal.2023}
P.~Uijtewaal, B.~Côté, T.~Foppen, J.~H. W.~d. Vries, S.~J. Woodings, P.~T.~S.
  Borman, S.~Lambert-Girard, F.~Therriault-Proulx, B.~W. Raaymakers, and M.~F.
  Fast,
\newblock {Performance of the HYPERSCINT scintillation dosimetry research
  platform for the 1.5 T MR-linac},
\newblock Physics in Medicine \& Biology  (2023).

\bibitem{Almond.1999}
P.~R. Almond, P.~J. Biggs, B.~M. Coursey, W.~F. Hanson, M.~S. Huq, R.~Nath, and
  D.~W.~O. Rogers,
\newblock {AAPM's TG‐51 protocol for clinical reference dosimetry of
  high‐energy photon and electron beams},
\newblock Medical Physics {\bf 26}, 1847--1870 (1999).

\bibitem{Azangwe.2014}
G.~Azangwe, P.~Grochowska, D.~Georg, J.~Izewska, J.~Hopfgartner, W.~Lechner,
  C.~E. Andersen, A.~R. Beierholm, J.~Helt‐Hansen, H.~Mizuno, A.~Fukumura,
  K.~Yajima, C.~Gouldstone, P.~Sharpe, A.~Meghzifene, and H.~Palmans,
\newblock Detector to detector corrections: a comprehensive experimental study
  of detector specific correction factors for beam output measurements for
  small radiotherapy beams,
\newblock Medical Physics {\bf 41}, 072103 (2014).

\bibitem{Morin.2013}
J.~Morin, D.~Béliveau‐Nadeau, E.~Chung, J.~Seuntjens, D.~Thériault,
  L.~Archambault, S.~Beddar, and L.~Beaulieu,
\newblock A comparative study of small field total scatter factors and dose
  profiles using plastic scintillation detectors and other stereotactic
  dosimeters: The case of the CyberKnife,
\newblock Medical Physics {\bf 40}, 011719 (2013).

\bibitem{Larraga-Gutierrez.2015}
J.~M. Lárraga-Gutiérrez, P.~Ballesteros-Zebadúa, M.~Rodríguez-Ponce, O.~A.
  García-Garduño, and O.~O. G. d.~l. Cruz,
\newblock {Properties of a commercial PTW-60019 synthetic diamond detector for
  the dosimetry of small radiotherapy beams},
\newblock Physics in Medicine \& Biology {\bf 60}, 905--924 (2015).

\bibitem{Mcewen2014}
M.~McEwen, L.~DeWerd, G.~Ibbott, D.~Followill, D.~W. Rogers, S.~Seltzer, and
  J.~Seuntjens,
\newblock Addendum to the AAPM's TG-51 protocol for clinical reference
  dosimetry of high-energy photon beams,
\newblock Medical physics {\bf 41}, 041501 (2014).

\bibitem{Duchaine.2022}
J.~Duchaine, D.~Markel, and H.~Bouchard,
\newblock {Efficient dose‐rate correction of silicon diode relative dose
  measurements},
\newblock Medical Physics {\bf 49}, 4056--4070 (2022).

\bibitem{Kry2012}
S.~F. Kry, R.~Popple, A.~Molineu, and D.~S. Followill,
\newblock Ion recombination correction factors () for Varian TrueBeam
  high-dose-rate therapy beams,
\newblock Journal of applied clinical medical physics {\bf 13}, 318--325
  (2012).

\bibitem{Bouchard.2009}
H.~Bouchard, J.~Seuntjens, J.~Carrier, and I.~Kawrakow,
\newblock {Ionization chamber gradient effects in nonstandard beam
  configurations},
\newblock Medical Physics {\bf 36}, 4654--4663 (2009).

\bibitem{Papaconstadopoulos.2014}
P.~Papaconstadopoulos, F.~Tessier, and J.~Seuntjens,
\newblock On the correction, perturbation and modification of small field
  detectors in relative dosimetry,
\newblock Physics in Medicine and Biology {\bf 59}, 5937--5952 (2014).

\bibitem{Lechner2020}
W.~Lechner, D.~Georg, and H.~Palmans,
\newblock An analytical formalism for the assessment of dose uncertainties due
  to positioning uncertainties,
\newblock Medical Physics {\bf 47}, 1357--1363 (2020).

\bibitem{Wulff.2008}
J.~Wulff, K.~Zink, and I.~Kawrakow,
\newblock {Efficiency improvements for ion chamber calculations in high energy
  photon beams: Efficiency improvements for ion chamber calculations},
\newblock Medical Physics {\bf 35}, 1328--1336 (2008).

\bibitem{Kawrakow.2000}
I.~Kawrakow,
\newblock {Accurate condensed history Monte Carlo simulation of electron
  transport. II. Application to ion chamber response simulations},
\newblock Medical Physics {\bf 27}, 499--513 (2000).

\bibitem{Kawrakow.20000d}
I.~Kawrakow,
\newblock {Accurate condensed history Monte Carlo simulation of electron
  transport. I. EGSnrc, the new EGS4 version},
\newblock Medical Physics {\bf 27}, 485--498 (2000).

\bibitem{rogers1995beam}
D.~Rogers, B.~Faddegon, G.~Ding, C.-M. Ma, J.~We, and T.~Mackie,
\newblock BEAM: a Monte Carlo code to simulate radiotherapy treatment units,
\newblock Medical physics {\bf 22}, 503--524 (1995).

\bibitem{Constantin.2011}
M.~Constantin, J.~Perl, T.~LoSasso, A.~Salop, D.~Whittum, A.~Narula, M.~Svatos,
  and P.~J. Keall,
\newblock {Modeling the TrueBeam linac using a CAD to Geant4 geometry
  implementation: Dose and IAEA-compliant phase space calculations: Varian
  TrueBeam Linac},
\newblock Medical Physics {\bf 38}, 4018--4024 (2011).

\bibitem{SciPy2020}
P.~Virtanen et~al.,
\newblock {{SciPy} 1.0: Fundamental Algorithms for Scientific Computing in
  Python},
\newblock Nature Methods {\bf 17}, 261--272 (2020).

\bibitem{Branch1999}
M.~Branch, T.~F. Coleman, and Y.~Li,
\newblock A Subspace, Interior, and Conjugate Gradient Method for Large-Scale
  Bound-Constrained Minimization Problems,
\newblock SIAM Journal on Scientific Computing {\bf 21}, 1--23 (1999).

\bibitem{Sauer.2007}
O.~A. Sauer and J.~Wilbert,
\newblock {Measurement of output factors for small photon beams},
\newblock Medical Physics {\bf 34}, 1983--1988 (2007).

\bibitem{Muir.2010}
B.~R. Muir and D.~W.~O. Rogers,
\newblock {Monte Carlo calculations of , the beam quality conversion factor},
\newblock Medical Physics {\bf 37}, 5939--5950 (2010).

\bibitem{Wulff.2010}
J.~Wulff, J.~T. Heverhagen, K.~Zink, and I.~Kawrakow,
\newblock {Investigation of systematic uncertainties in Monte Carlo-calculated
  beam quality correction factors},
\newblock Physics in Medicine \& Biology {\bf 55}, 4481--4493 (2010).

\bibitem{Francescon.2014}
P.~Francescon, W.~Kilby, and N.~Satariano,
\newblock Monte Carlo simulated correction factors for output factor
  measurement with the CyberKnife system—results for new detectors and
  correction factor dependence on measurement distance and detector
  orientation,
\newblock Physics in Medicine and Biology {\bf 59}, N11--N17 (2014).

\bibitem{Welch1947}
B.~L. Welch,
\newblock The generalization of "Student's" problem when several different
  population variances are involved",
\newblock Biometrika {\bf 34}, 28--35 (1947).

\bibitem{Looe2018}
H.~K. Looe, I.~B{\"u}sing, T.~Tekin, A.~Brant, B.~Delfs, D.~Poppinga, and
  B.~Poppe,
\newblock The polarity effect of compact ionization chambers used for small
  field dosimetry,
\newblock Medical Physics {\bf 45}, 5608--5621 (2018).

\bibitem{gul2020}
A.~Gul, S.~Fukuda, H.~Mizuno, N.~Taku, M.~B. Kakakhel, and S.~M. Mirza,
\newblock Feasibility study of using stereotactic field diode for field output
  factors measurement and evaluating three new detectors for small field
  relative dosimetry of 6 and 10 MV photon beams,
\newblock Journal of Applied Clinical Medical Physics {\bf 21}, 23--36 (2020).

\bibitem{birks1956scintillation}
J.~Birks and F.~Brooks,
\newblock Scintillation response of anthracene to 6-30 keV photoelectrons,
\newblock Proceedings of the Physical Society. Section B {\bf 69}, 721 (1956).

\bibitem{Lechner.2022}
W.~Lechner, R.~Alfonso, M.~Arib, M.~S. Huq, A.~Ismail, R.~Kinhikar, J.~M.
  Lárraga-Gutiérrez, K.~R. Mani, N.~Maphumulo, O.~A. Sauer, S.~Shoeir,
  S.~Suriyapee, and K.~Christaki,
\newblock A multi-institutional evaluation of small field output factor
  determination following the recommendations of {IAEA}/{AAPM} {TRS}-483,
\newblock Medical Physics {\bf 49}, 5537--5550 (2022).

\bibitem{Mateus2023}
D.~Mateus, C.~Greco, and L.~Peralta,
\newblock Field Output Correction Factors of Small Static Field for {{IBA}}
  Razor Nanochamber,
\newblock Biomedical Physics \& Engineering Express {\bf 10}, 015004 (2024).

\bibitem{Das.2018}
I.~J. Das and P.~Francescon,
\newblock {Comments on the TRS‐483 protocol on small field dosimetry},
\newblock Medical Physics {\bf 45}, 5666--5668 (2018).

\bibitem{Palmans2018}
H.~Palmans, P.~Andreo, M.~S. Huq, J.~Seuntjens, K.~E. Christaki, and
  A.~Meghzifene,
\newblock Reply to “{{Comments}} on the {{{\textsc{TRS}}}} ‐483
  {{Protocol}} on {{Small}} Field {{Dosimetry}}” [{{Med}}. {{Phys}}. 45(12),
  5666–5668 (2018)],
\newblock Medical Physics {\bf 45}, 5669--5671 (2018).

\bibitem{Ringholz2023}
J.~Ringholz, O.~A. Sauer, and S.~Wegener,
\newblock Small Field Output Correction Factors at 18 {{MV}},
\newblock Medical Physics {\bf 50}, 7177--7191 (2023).

\bibitem{Looe.2019}
H.~K. Looe, D.~Poppinga, R.~Kranzer, I.~Büsing, T.~Tekin, A.~Ulrichs,
  B.~Delfs, D.~Vogt, J.~Würfel, and B.~Poppe,
\newblock The role of radiation-induced charge imbalance on the dose-response
  of a commercial synthetic diamond detector in small ﬁeld dosimetry,
\newblock Medical Physics {\bf 46}, 2752--2759 (2019).

\bibitem{Girardi.2018P179}
A.~Girardi, T.~Gevaert, and M.~De~Ridder,
\newblock [P179] Small field correction factors determination for the IBA razor
  nano chamber and the IBA razor chamber,
\newblock Physica Medica: European Journal of Medical Physics {\bf 52},
  151--152 (2018).

\end{thebibliography}
